\title{s3d-paper}
\documentclass{bmcart}

\usepackage{amsthm,amsmath}
\RequirePackage{natbib}
\RequirePackage{hyperref}
\usepackage[utf8]{inputenc} 

\usepackage[capitalise,noabbrev]{cleveref}
\usepackage{booktabs}
\usepackage{mathtools}
\usepackage{color}
\usepackage{multirow}
\usepackage{tabularx}
\usepackage{array}
\newcolumntype{L}[1]{>{\raggedright\let\newline\\\arraybackslash\hspace{0pt}}m{#1}}
\newcolumntype{C}[1]{>{\centering\let\newline\\\arraybackslash\hspace{0pt}}m{#1}}
\newcolumntype{R}[1]{>{\raggedleft\let\newline\\\arraybackslash\hspace{0pt}}m{#1}}

\usepackage{graphicx}
\usepackage{url}
\usepackage{subcaption}

\usepackage{xcite}
\usepackage{xr}
\makeatletter
\newcommand*{\addFileDependency}[1]{
  \typeout{(#1)}
  \@addtofilelist{#1}
  \IfFileExists{#1}{}{\typeout{No file #1.}}
}
\makeatother

\newcommand*{\myexternaldocument}[1]{%
    \externaldocument{#1}%
    \addFileDependency{#1.tex}%
    \addFileDependency{#1.aux}%
}

\myexternaldocument{supplementary}

\startlocaldefs
\endlocaldefs

\begin{document}

\begin{frontmatter}

\begin{fmbox}
\dochead{Research}


\title{Predicting and Explaining Behavioral Data with Structured Feature Space Decomposition}


\author[
   addressref={aff1},                   
   email={peter.g.fennell@gmail.com }   
]{\inits{PGF}\fnm{Peter G} \snm{Fennell}}
\author[
   addressref={aff1,aff2},
   email={zhiya-zuo@uiowa.edu }
]{\inits{ZZ}\fnm{Zhiya} \snm{Zuo}}
\author[
   addressref={aff1},
   corref={aff1},                       
   noteref={n1},                        
   email={lerman@isi.edu}
]{\inits{KL}\fnm{Kristina} \snm{Lerman}}


\address[id=aff1]{
  \orgname{USC Information Sciences Institute}, 
  \street{4676 Admiralty Way},                     %
  \city{Marina del Rey, CA},                              
  \cny{USA}                                    
}
\address[id=aff2]{%
  \orgname{University of Iowa},
  \city{Iowa City, IA},
  \cny{USA}
}


\begin{artnotes}
\note[id=n1]{Equal contributor} 
\end{artnotes}

\end{fmbox}


\begin{abstractbox}

\begin{abstract} 
Modeling human behavioral data is challenging due to its scale, sparseness (few observations per individual), heterogeneity (differently behaving individuals), and class imbalance (few observations of the outcome of interest). An additional challenge is learning an interpretable model that not only accurately predicts outcomes, but also identifies important factors associated with a given behavior. To address these challenges, we describe a statistical approach to modeling behavioral data called the structured sum-of-squares decomposition (S3D). The algorithm, which is inspired by decision trees,  selects important features that collectively explain the variation of the outcome, quantifies correlations between the features, and partitions the subspace of important features into smaller, more homogeneous blocks that correspond to similarly-behaving subgroups within the population. This partitioned subspace allows us to predict and analyze the behavior of the outcome variable both statistically and visually, giving a medium to examine the effect of various features and to create explainable predictions.
We apply S3D to learn models of online activity from large-scale data collected from diverse sites, such as Stack Exchange, Khan Academy, Twitter, Duolingo, and Digg. We show that S3D creates parsimonious models that can predict outcomes in the held-out data at levels comparable to state-of-the-art approaches, but in addition, produces interpretable models that provide insights into behaviors. This is important for informing strategies aimed at changing behavior, designing social systems, but also for explaining predictions, a critical step towards minimizing algorithmic bias.
\end{abstract}


\begin{keyword}
\kwd{computational social science}
\kwd{empirical studies}
\kwd{online social networks}
\kwd{human behavior}
\kwd{feature selection}
\end{keyword}


\end{abstractbox}
%

\end{frontmatter}



\section{Introduction}
Explanation and prediction  are complementary goals of computational social science~\cite{Watts2018prediction}. The former focuses on   identifying factors  that explain human behavior, for example, by using regression to estimate parameters of theoretically-motivated models from data. Insights gleaned from such interpretable models have been used to inform the design of social platforms~\cite{Settles2016} and  intervention strategies that steer human behavior in a desired direction~\cite{nudge}. In recent years, prediction  has become the de-facto standard for evaluating learned models of social data~\cite{Lazer2009}. This trend, partly driven by the dramatic growth of  behavioral data and the success of machine learning algorithms, such as decision trees and support vector machines, emphasizes ability to accurately predict unseen cases (out-of-sample or held out data) over learning interpretable models~\cite{Lipton2018,HofmanSharma2017}.

Motivated by the joint goals of prediction and explanation, we propose Structured Sum-of-Squares Decomposition (S3D) algorithm, a method for learning interpretable statistical models of behavioral data. The algorithm, a variant of regression trees~\cite{Breiman1984},  builds a single tree-like structure that is both highly \textit{interpretable} and can be used for out-of-sample \textit{prediction}. In addition, the learned models can be used to \textit{visualize} data.
S3D is a mathematically principled method that addresses the computational challenges associated with mining behavioral data. Such data is usually massive, containing records of many individuals, each with a large number of, potentially highly correlated, features. However, the data is also  sparse (with only a few observations available per individual) and unbalanced (few examples of the behavior within each class). Yet another challenge is heterogeneity:  data is composed of subgroups that vary widely in their behavior. For example, the vast bulk of social media users have very few followers and post a few messages, but a few users have millions of followers or are extremely prolific posters. Ignoring heterogeneity could lead analysis to wrong conclusions due to statistical paradoxes~\cite{Blyth1972,Alipourfard2018}.

S3D works as follows: given a set of features $\{X_i\}_{i=1}^M$ and an outcome variable $Y$, S3D identifies a subset of $m$ important features that are orthogonal in their relationship with the outcome and collectively explain the largest amount of the variation in the outcome. In addition to these \emph{selected} features, S3D identifies correlations between all features, thus providing important insights into the effects of features that were not selected by the model. Similar to regression trees and other decision trees, the S3D algorithm recursively partitions the $m$-dimensional space defined by the selected features into smaller, more homogeneous subgroups or bins, where the outcome variable exhibits little variation within each bin but significant variation between bins; however, it does so in a structured way, by minimizing variation in the outcome \textit{conditioned} on the existing partition.  The decomposition effectively creates an approximation of the (potentially non-linear) functional relationship between $Y$ and the features, while the structured nature of the decomposition gives the model interpretability and also reduces overfitting.
The resulting model is parsimonious and, despite its low complexity, is a highly performant predictive tool.

To demonstrate the utility of the proposed method, we apply it to model a variety of datasets, from  benchmarks to large-scale heterogeneous behavioral data that we collect from social platforms, including Twitter, Digg, Khan Academy, Duolingo, and Stack Exchange. Across datasets, performance of S3D is competitive to existing state-of-the-art methods on both classification and regression tasks, while it also offers several advantages. We highlight these advantages by showing how S3D reveals the important factors in explaining and predicting behaviors, such as experience, skill, and answer complexity when analyzing performance on the question answering site Stack Exchange or essential nodal attributes, such as activity and rate of receiving information on the social networks Digg and Twitter. Qualitatively, S3D visualizes the relationship between the outcome and features, and quantifies their importance via prediction task. Surprisingly, despite high heterogeneity of these relationships in many datasets, just a few important features identified by S3D can predict held-out data with remarkable accuracy.

Machine learning, data science, and social science communities have proposed different solutions to learning models of data. Popular among these are regression methods, such as linear and logistic regression. Lasso~\cite{Hastie2009} and elastic net~\cite{Zou2005} improve on standard regression, using regularization to reduce the dimensionality of the feature space and to prevent overfitting. Decision trees and their regression tree variants that inspired S3D, such as CART~\cite{Breiman1984}, BART~\cite{Chipman2010}, and MARS~\cite{Friedman1991},  partition data into non overlapping subsets to minimize the variance of response variable, or some other cost function, within these subsets. Ensemble tree-based methods, such as random forests~\cite{Breiman2001}, learn a model as a sum or ensemble of individual decision trees. However, while these approaches address one set of challenges, they often trip over the remaining ones. Regression models (e.g., ridge, lasso, elastic net) are limited by their specified functional form, and cannot capture relationships in data that do not adhere to this form.
Decision trees, including random forests, on the other hand, are very effective at capturing non-linear and unbalaced data, but have limited interpretability. While they offer a measure of feature importance, the relationship between the outcome and features is not easily analyzable: explaining the predictions requires navigating the depths of many trees---with features potentially repeating at different levels---a herculean task at best. Our method, on the other hand, is fully transparent and allows for visualizing and explaining why predictions were made. S3D has lower complexity than random forests, which typically consist of dozens of trees; indeed S3D also has the advantage of having only two tunable parameters, as opposed to twelve for a typical random forest. Finally, S3D adds structure to the feature importance offered by random forests, showing which subset of features are sufficient for explaining variation in outcomes, as well constructing feature networks which show correlation between features and where the variance from the other features is absorbed.

S3D learns a low complexity model with only two hyperparameters. Its ability to capture non-linear and heterogeneous behavior, while remaining interpretable, makes it a valuable computational tool for the analysis of behavioral data.

\section{Method}
\label{sec:methods}

We present the Structured Sum-of-Squares Decomposition algorithm (S3D), a variant of regression trees~\cite{Breiman1984,Chipman2010,Friedman1991}, that takes as input a set of features $\mathbf{X}=\{X_i\}_{i=1}^M$ and an outcome variable $Y$ and selects a smaller set of $m$ important features that collectively best explain the outcome. The method partitions the values of each important feature $X^S_i$ to decompose the $m$-dimensional selected feature space 
into smaller non-overlapping blocks, 
such that $Y$ exhibits significant variation between blocks but little variation within each block.
These blocks 
allow us to approximate the functional relationship $Y=f(\mathbf{X})$ as a multidimensional step function over all blocks in the $m$-dimensional selected feature space, thus capturing non-linear relationships between features and the outcome.

Our method chooses features recursively in a forward selection manner, so that features chosen at each step explain most of the variation in $Y$ \emph{conditioned} on the features chosen at the previous steps. Features that are correlated will explain much of the same variation in $Y$, and our approach of successively choosing features based on how much remaining variation in $Y$ they explain results in a set $\mathbf{X}^S = \{X^S_i\}_{i=1}^m$ of important features that are orthogonal in their relationships with $Y$.
By decomposing data recursively, we create a parsimonious model that quantifies relationships not only between the features $X_i$ and the outcome variable $Y$, but also among the features themselves. We show that 
our model is able to achieve performance comparable to state-of-the-art machine learning algorithms on prediction tasks, while offering advantages over those methods: our algorithm uses only two tuning parameters, can represent non-linear relationships between variables, and creates an interpretable model that is amenable to analysis and produces insights into behavior that merely predictive models do not give.

\subsection{Structured Feature Space Decomposition}
\label{sec:learning}

A key concept used to describe variation in observations $\{y_i\}_{i=1}^N$ of a random variable $Y$ is the total sum of squares $SST$, which is defined as
	$SST = \sum_{i=1}^N(y_i - \bar{y})^2,$
where $\bar{y} = \sum_{i=1}^Ny_i/N$ is the sample mean of the observations. The total sum of squares is intrinsically related to variation in $Y$; indeed the sample variance $\hat{\sigma}^2$ of $Y$ can be directly obtained from this quantity as $\hat{\sigma}^2 = SST/(N-1)$.

Given a feature $X_j$, one method of quantifying how much variation in $Y$ can be explained by $X_j$ is as follows: (\textit{1}) partition $X_j$ into a collection $P_{X_j}$ of non-overlapping bins, (\textit{2}) compute the number of data points $N_p$ and the average value $\bar{y}_p$ of $Y$ in each bin $p\in P_{X_j}$, and (\textit{3}) decompose the total sum of squares of $Y$ as 
\begin{equation}
	\sum_{i=1}^N(y_i - \bar{y})^2 = \sum_{p\in P_{X_j}}N_p(\bar{y}_p - \bar{y})^2 + \sum_{p\in P_{X_j}}\sum_{i=1}^{N_p}(y_{p,i} - \bar{y}_p)^2,
	\label{eq:SST_decomposition}
\end{equation}
where $y_{p,i}$ here is the $i$'th data point in bin $p$. The first sum on the right hand side of Eq.~\eqref{eq:SST_decomposition} is the explained sum of squares (or sum of squares between groups), a weighted average of squared differences between global ($\bar{y}$) and local ($\bar{y}_p$) averages that measures how much $Y$ varies between different bins $p$ of $X_j$. The second sum is the residual sum of squares (or sum of squares within groups), which measures how much variation in $Y$ remains within each bin $p$. The $R^2$ coefficient of determination is then the proportion of the explained sum of squares to the total sum of squares, given by
\begin{equation}
	R^2 = \frac{\sum_{p\in P_{X_j}}N_p(\bar{y}_p - \bar{y})^2}{SST}.
	\label{eq:R2}
\end{equation}
The $R^2$ measure takes values between zero and one, with large values of $R^2$ indicating a larger proportion of the variation of $Y$ explained by $X_j$. This method of approximating the functional relationship between $Y$ and $X_j$ as a step function with bins, or groups $P_{X_j}$ and corresponding values $\bar{y}_p$, allows us to quantify the variation in $Y$ explained by $X_j$ through the $R^2$ of the corresponding step function as given by Eq.~\eqref{eq:R2}.

\subsubsection{Partitioning Values of a Feature}

We now introduce a method to systematically \emph{learn} the partition $P_{X_j}$ of the feature $X_j$ which will be central to our algorithm. Given the data, we can split the domain of the feature $X_j$ at the value $s$ into two bins: $X_j \leq s$ and $X_j > s$. From Eq.~\eqref{eq:R2}, we see that the proportion of variation in $Y$ explained by such a split is
\begin{equation}
	R^2(s;X_j) =\frac{N_{X_j\leq s}(\bar{y}_{X_j\leq s} - \bar{y})^2 + N_{X_j > s}(\bar{y}_{X_j > s} - \bar{y})^2}{SST},
	\label{eq:R2bin1}
\end{equation}
where $N_{X_j\leq s}$ and $\bar{y}_{X_j\leq s}$ are the number of data points and average value of $Y$ in the bin $X_j \leq s$, and vice versa for $N_{X_j > s}$ and $\bar{y}_{X_j > s}$. $R^2(s;X_j)$ can be computed for each possible value of $s$ in the domain of $X_j$, and we can choose the optimal split $s_1$ as the split $s$ that maximizes $R^2(s;X_j)$ of Eq.~\eqref{eq:R2bin1}. Choosing $s_1$, and partitioning the domain of $X_j$ into $P_{X_j} = \{[\min(X_j),s_1],(s_1,\max(X_j)]\}$, we can again find the next best split $s_2$ to optimize the improvement in $R^2$. In general, having chosen $n$ splits $\{s_u\}_{u=1}^n$ with a resulting partition $P_{X_j}$ of $n+1$ bins, the next best split $s_{n+1}$ can be chosen as the split $s$ that maximizes the improvement in $R^2$ as given by
\begin{multline}
	\Delta R^2(s|P_{X_j}; X_j) = \frac{1}{SST} \Big( N_{p(s)|X_j \leq s}(\bar{y}_{p(s)|X_j \leq s})^2 \\ + N_{p(s)|X_j > s}(\bar{y}_{p(s)|X_j > s})^2 - N_{p(s)}(\bar{y}_{p(s)})^2 \Big)
	\label{eq:binning}
\end{multline}
where $p(s)$ in Eq.~\eqref{eq:binning} is the bin in $P_{X_j}$ that contains the point $s$ and $p(s)|X_j \leq s$ (resp. $p(s)|X_j > s$) is the restriction of that bin to points $X_j \leq s$ (resp. $X_j > s$). In this manner, we recursively split the domain of $X_j$ to create a partition of the feature.

However, splitting in this manner can continue indefinitely, resulting in a model that is too fine-grained and thus overfits the data. To prevent overfitting, we need a stopping criterion. To this end we introduce a loss function $L(P_{X_j})$ that penalizes the size $|P_{X_j}|$ of the partition, i.e., the number of bins:
\begin{equation}
	L(P_{X_j}) = 1-R^2(P_{X_j}) + \lambda |P_{X_j}|.
	\label{eq:lossfunction}
\end{equation}
The parameter $\lambda$ controls how fine-grained the bins are: smaller values of $\lambda$ allow for more finer bins, and vice versa. The loss function of Eq.~\eqref{eq:lossfunction} reaches a minimum when the change in $R^2(P_{X_j})$ from adding an extra split to $P_{X_j}$ is less than $\lambda$---at this point we stop splitting and return the partition $P_{X_j}$.

Having formed the partition $P_{X_j}$ of $X_j$ with splits $\{s_u\}_{u=1}^n$, the total score $R^2(X_j)$ can be calculated from Eq.~\eqref{eq:R2}, or by summing $R^2(s_1;X_j)$ from Eq.~\eqref{eq:R2bin1} along with the $\Delta R^2$ terms in Eq.\eqref{eq:binning} for each of the splits $\{s_u\}_{u=2}^n$. Completing this procedure for all features gives a measure of how much variation in $Y$ each feature alone explains, and ranking these features allows us to choose the most important feature $X_{C_1}$ that explains the largest amount of the variation in $Y$.

\subsubsection{Selecting Additional Features}

After choosing the most important feature, we search the rest of the features for one that explains most of the remaining variance in $Y$, then the third feature, and so on. Here, we describe the procedure for finding the next best feature having already chosen $l$ features $\mathbf{X}^S = \{X^S_1,\dots,X^S_l\}$ with a corresponding partition $\mathcal{P}^S = P^S_1\times\dots\times P^S_{l}$, where $\times$ here is the cartesian product. In this case, a total $R^2(\mathcal{P}^S) = \sum_{p\in \mathcal{P}^S}N_p(\bar{y}_p - \bar{y})^2/SST$ of the variation in $Y$ has been explained, and we now look for the feature that best explains the remaining variation $1-R^2(\mathcal{P}^S)$.

Given a remaining feature $X_j$, we partition the domain of $X_j$ similarly to how we partitioned it when choosing the first feature. The first split $s_1$ of $X_j$ is chosen as the value $s$ that maximizes the improvement in $R^2$, given by
\begin{multline}
	\Delta R^2(s|\mathcal{P}^S; X_j) = \frac{1}{SST}\sum_{p\in \mathcal{P}^S} \Big( N_{p|X_j\leq s}(\bar{y}_{p|X_j\leq s})^2 \\ + N_{p|X_j>s}(\bar{y}_{X_j>s})^2 - N_{p}(\bar{y}_{p})^2\Big),
\end{multline}
where $p|X_j\leq s$ (resp. $p|X_j > s$) is the set of data points in $p\in \mathcal{P}^S$ for which $X_j\leq s$ (resp. $X_j>s$). In general, given $n$ splits and a corresponding partition $P_{X_j}$ of $X_j$,  the $n+1$'st split is chosen as the value $s$ that maximizes
\begin{multline}
	\Delta R^2(s|P_{X_j},\mathcal{P}^S; X_j) = \frac{1}{SST}\sum_{p\in \mathcal{P}^S\times P_{X_j}|s\in p} \Big( N_{p|X_j\leq s}(\bar{y}_{p|X_j\leq s})^2 \\ + N_{p|X_j>s}(\bar{y}_{X_j>s})^2 - N_{p}(\bar{y}_{p})^2\Big),
	\label{eq:split_sub}
\end{multline}
with the sum in Eq.~\eqref{eq:split_sub} taken over all elements $p$ of $\mathcal{P}^S\times P_{X_j}$ that contain the point $s$. The loss function in the general setting is
\begin{equation}
	L(P_{X_j}|\mathcal{P}^S) = \frac{1-R^2(\mathcal{P}^S\times P_{X_j})}{1-R^2(\mathcal{P}^S)} + \lambda |P_{X_j}|,
	\label{eq:lossfunction_general}
\end{equation}
where the denominator of the fractional first term in Eq.~\eqref{eq:lossfunction_general} normalizes this term to be between zero and one, is the case in Eq.~\eqref{eq:lossfunction}. This normalization is necessary because as we progress through the algorithm, subsequent features may explain less of the variance of $Y$ (as features are chosen hierarchically), and so changes in $1-R^2(\mathcal{P}^S\times P_{X_j})$ by splitting $X_j$ are smaller. The effect of this would be a coarser binning of the feature, and so the normalization ensures that this is not the case and that the feature is binned consistently at each stage of the algorithm. 
Again, the partition $P_{X_j}$ of $X_j$ is chosen that minimizes the loss function of Eq.~\eqref{eq:lossfunction_general}, and the $R^2$ improvement is calculated for this feature as
\begin{equation}
	\Delta R^2(P_{X_j}|\mathcal{P}^S) = R^2(\mathcal{P}^S\times P_{X_j}) - R^2(\mathcal{P}^S).
	\label{eq:bestR2}
\end{equation}
This procedure is repeated for all remaining features $X_j$ to select the feature $X_{l+1}$ with the maximal $R^2$ improvement.

This process of binning features, calculating their improvement $\Delta R^2$ and choosing the one with the largest improvement continues until no further variation in $Y$ can be explained or until an alternative stopping condition (such as a pre-specified maximum number of steps) is met. Our algorithm learns a hierarchy of important features that explain the variation in $Y$ and a decomposition $\mathcal{P}^S$ with corresponding values $\{\bar{y}_p\}_{p\in \mathcal{P}^S}$ approximating the functional relationship between the outcome and the features. Note that, when binning a feature, once the vectors $\{\{N_{p|X_j\leq s}\}_{p\in \mathcal{P}^S}\}_s$ and $\{\{\bar{y}_{p|X_j\leq s}\}_{p\in \mathcal{P}^S}\}_s$ have been constructed (an operation that takes $\mathcal{O}(N)$ time), the binning is independent of $N$, and instead depends on the number of unique values of the feature (as required to calculate the optimal splits $s$ at each step).
The fact that S3D scales linearly with the number of data points $N$
allows us to apply the algorithm to large datasets, such as the Twitter and Digg (see Section~\ref{sec:results}).

%

\subsubsection{Hyperparameters}
The S3D model has two hyperparameters: (\textit{1}) $\lambda$ that controls granularity of 
feature binning; (\textit{2}) $k$ that specifies the number of features to use for prediction. Both are important to prevent overfitting --- as left unrestricted, the algorithm can learn too fine-grained a model that fails to generalize to unseen data. We note that it is possible to stop early in the training phase by restricting the maximum number of features to select. In other words, the algorithm can stop when the number of selected features reaches a predefined value. Nonetheless, it is recommended \emph{not} to lay any limit during the training phase but rather tune $k$ in the validation step.

To tune the hyperparameters, we train S3D for various values of $\lambda$, in each case letting the algorithm continue until there is no further improvement in $R^2$.  This results in a model with $m_{\lambda}$ selected features and partition $\mathcal{P}^{S_{\lambda}}=\{P^{S_{\lambda}}_1,...,P^{S_{\lambda}}_{m_{\lambda}}\}$. Then, for 
$k \in \big[1, m_{\lambda}\big]$, we evaluate the predictive performance of the model using only the top 
$k$ selected features and the sub-partition $\{P^{S_{\lambda}}_1,...,
P^{S_{\lambda}}_{k}\}$. Performance is measured on held-out tuning data using a specified metric. 
The optimal hyperparameters $(\lambda,k)$  are those that achieve the best performance on held-out data.

\subsection{Applications of the Learned Model}

Given a dataset, S3D learns an ordered set of important, orthogonal features $\mathbf{X}^S$, a partitioning $\mathcal{P}^S$ of the selected feature space with corresponding $\bar{y}_p$ and $N_p$ values for each block $p\in \mathcal{P}^S$, and $\Delta R^2$ for each remaining variable at each step of the algorithm. This decomposition serves as a parsimonious model of data and can be used for feature selection, feature correlation, prediction and analysis as described below.

\subsubsection{Feature Selection and Correlations}
\label{sec:feat_selec_corr}
The ordered set $\mathbf{X}^S$ of important, orthogonal features allows us to quantify feature importance in heterogeneous behavioral data. The top-ranked features explain the largest amount of variation in the outcome variable, while each successive feature explains most of the remaining variation that is not explained by the features that were already selected.

Aside from the selected features $\mathbf{X}^S$, S3D provides insights into features that are \emph{not} selected by the algorithm, quantifying variation that they explain in the outcome variable that is made redundant through the selected variables. This is calculated in the following manner. At a given step $l$ of the algorithm, feature $X^S_l$ is selected as the best feature with an $R^2$ improvement of $\Delta R^2(P_{X^S_l}|\mathcal{P}^{S(l-1)})$ (Eq.~\eqref{eq:bestR2}), where $\mathcal{P}^{S(l-1)}$ is the partition prior to step $l$. Meanwhile, a different remaining feature $X_j$ has an $R^2$ improvement of $\Delta R^2(P_{X_j}|\mathcal{P}^{S(l-1)})$. At the next stage of the algorithm, given $X^S_l$ has been selected, $X_j$ will have an $R^2$ improvement of $\Delta R^2(P_{X_j}|\mathcal{P}^{S(l-1)}\times P_{X^S_l})$, and thus the variation in $X_j$ that is made redundant through the selection of $X^S_l$ is the difference between these two $\Delta R^2$:
\begin{equation}
  a_{X_j,X^S_l} = \Delta R^2\big(P_{X_j}\big|\mathcal{P}^{S(l-1)}\big) - \Delta R^2\big(P_{X_j}\big|\mathcal{P}^{S(l-1)}\times P_{X^S_l}\big).
  \label{eq:net_coeffs}
\end{equation}
The coefficients $a_{X_j,X^S_i}$ facilitate our analysis of the effect of unselected features on the outcome variable, and we implement them as weights in a feature network that is weighted and directed. This network gives a tool for analysis---unselected features that otherwise can explain much of the variation in the outcome will have heavy links, and the selected features $X^S_l$ to which these links point reveal correlations and through which selected features the unselected feature is made redundant.

\subsubsection{Prediction}
\label{sec:prediction}
The learned model can be used as a predictive tool for both discrete and continuous-valued outcome variables. Given input data $\mathbf{x}$, the model predicts the expected value $\hat{\mu}$ of $Y$ as $\hat{\mu}|\mathbf{x} = \bar{y}_{p(\mathbf{x})}$, where $p(\mathbf{x})$ is the block in decomposition $\mathcal{P}^S$ to which $\mathbf{x}$ belongs. For continuous-valued outcome variables, this predicted expected value $\hat{\mu}$ will be the prediction of the outcome of $Y$, i.e., $\hat{y}|\mathbf{x} = \bar{y}_{p(\mathbf{x})}$. For discrete-valued outcomes, the expected value has to be thresholded to predict an outcome class. For binary outcomes  $Y\in \{0,1\}$, $\hat{\mu}|\mathbf{x} = \bar{y}_{p(\mathbf{x})}$ is the maximum likelihood estimate of the probability of the outcome $Y=1$ in the block $p(\mathbf{x})$, and thus our model specifies that the outcome $Y=1|\mathbf{x}$ will occur with probability $\bar{y}_{p(\mathbf{x})}$. By choosing an appropriate \emph{discrimination threshold} $\theta$, our model then makes the prediction $\hat{y}$ as
\begin{equation}
	\hat{y}|\mathbf{x} =
	\begin{cases}
		1 & \bar{y}_{p(\mathbf{x})} \geq \theta \\
		0 & \bar{y}_{p(\mathbf{x})} < \theta.
	\end{cases}
\end{equation}

\paragraph{Unbalanced Data.}
Two of the datasets that we study, Digg and Twitter, are highly unbalanced---their  outcome variables (whether the user $adopts$ a meme) are binary, and the proportions of positive outcomes in the data are 0.0025 and 0.0007 respectively. Using  the standard discrimination  threshold $\theta=0.5$ results in predicting an insufficient number of ones. To address this issue, we choose the discrimination threshold based on the training data, picking the largest value $\theta=\theta\prime$, such that the number of predicted positive examples in the training data is greater than or equal to the actual number of positive examples in the training set. This threshold is then used for prediction on the held-out tuning data, as well as on the test data. Note that, for these two datasets, we also alter the discrimination threshold for the regression and random forest models in the same manner.

\subsubsection{Analysis and Model Interpretation}

One of the more novel contributions of S3D  is its potential for model exploration. By selecting features 
sequentially, we create a model where typically lower amounts of variation are explained at successive levels, and so a visual analysis of the first few important dimensions of the model allows us to understand the effects of the important features on data and predictions. The expectations $\bar{y}_p$ for each block $p\in \mathcal{P}^S$ facilitate this exploration, approximating the relationship $Y=f(\mathbf{X})$ between outcome and features. Furthermore, for binary data, the predicted outcomes obtained by thresholding the expectations show how predictions change as a function of the features, which allows for explaining predictions and visually exploring the data. 

\subsection{Comparison to State-of-the-art}
\label{sec:state-of-the-art}
We compare S3D to logistic regression (Lasso and elastic net~\cite{Zou2005}), random forests \cite{Breiman2001}, and support vector machines with linear kernel (linear SVM)~\cite{Boser1992}, using 5-fold cross validation (CV; \cref{supp-sec:cv}). These algorithms are ideal benchmarks for S3D, with the ability to provide feature importance scores and therefore interpretability to the trained models. Further, we emphasize that S3D can  produce sparse models that can do both feature selection and prediction. As shown in \cref{sec:results_benchmark_comparison}, S3D performs similarly to logistic regression in both classification and regression (\cref{fig:classification_comparison,fig:regression_comparison}), but uses fewer features (\cref{fig:num_features}). Finally, both random forests \cite{FernandezandManuel2014} and linear SVM \cite{ChangHsiehChange2010} are proven to be adequately powerful in prediction tasks while being relatively simple to train.

We partition data into five equal size folds,\footnote{For classification tasks, each fold is made by preserving the ratio of samples in each class (i.e., stratified.)} each of which is rotated as a hold-out set for testing. For classification (except random forests), we standardize feature values by centering and scaling variance to one. For regression, both features and target values are standardized. Standardization of test set is based on training data.
In each run, we train and tune the models on four folds, where three are used for training and one for validation. Finally we evaluate the performance of the tuned models on the remaining test fold. The final evaluation is therefore the average performance across each of the five folds. In other words, we tune the hyperparameters with 4-fold CV (hereafter \textit{inner CV}) and evaluate the performance of the optimal models with 5-fold CV (hereafter \textit{outer CV}.)
For classification tasks, we evaluate performance using (\textit{1}) accuracy (i.e., the percentage of correctly classified data points), (\textit{2}) $F1$ score (i.e., the harmonic mean of precision, the percentage of predicted ones that are correctly classified, and recall, the percentage of actual ones that are correctly classified), and (\textit{3}) area under the curve (AUC.) For regression tasks, we employ (\textit{1}) root mean squared error (RMSE),
(\textit{2}) mean absolute error (MAE-Mean),
(\textit{3}) median absolute error (MAE-Median).
Note that for classification tasks, higher values of the metrics imply better performance. For regression tasks, lower values of the metrics imply better performance. During the inner CV phase (i.e., hyperparameter tuning), we optimize classification performance on AUC scores and regression on RMSE scores.

For Lasso regression we tune the strength of $l1$ regularization. For elastic net regression, we tune the strengths of both $l1$ and $l2$ regularizations. For random forests, we tune  (\textit{1}) the number of features to randomly select for each decision tree, (\textit{2}) the minimum number of samples required to make a prediction, (\textit{3}) whether or not to bootstrap when sampling data for each decision tree, and (\textit{4}) criterion of the quality of a split (choices are Gini impurity or information gain for classification; MAE or MSE for regression.)
For linear SVM, we tune the penalty parameter for regularization.

Finally, we compare performance of models of similar complexity. Since S3D makes predictions using a small number of features, we also retrain the random forest on a similarly small number of features (although,  random forest is more complex, since it is an ensemble method). First, we use random forest to rank all the features and retrain the model on the top-$k$ (and top-$2k$) ranked features, where $k$ is the number of important features selected by S3D. This enables us to compare the  performance of models that use similar number of features. In addition, we investigate the effects of using S3D for \emph{supervised feature selection}. Specifically, we train random forests using only features chosen by S3D (i.e., $k$ features). 

\section{Results}
\label{sec:results}

We apply S3D\footnote{S3D is available as C++ code and Python wrapper: see \url{https://github.com/peterfennell/S3D}.} to benchmark datasets from the UCI Machine Learning Repository \cite{Dua:2017} and Lu\a'is Torgo's personal website. \footnote{\url{https://www.dcc.fc.up.pt/~ltorgo/Regression/DataSets.html}} In addition, we include five large-scale behavioral datasets, as described in the following paragraphs. \cref{tab:dataset} lists all \emph{18} datasets used in both classification and regression tasks, along with their statistics. See \cref{supp-sec:data} for more detailed description on data preprocessing.

Behavioral data came from various social platforms:
\begin{itemize}

\item \emph{Stack Exchange}.
The Q\&A platform Stack Exchange enables users to ask and answer questions. Askers can also \emph{accept} one of the answers as the best answer. This enables us to measure answerer performance by whether their answer was accepted as the best answer or not. The data we analyze includes 
a random sample of all answers posted on Stack Exchange from 08/2009 until 09/2014 
that preserves the class distribution. Each record corresponds to an answer and contains a binary outcome variable $Y\in\{0,1\}$ (one indicates the answer was accepted, and zero otherwise), along with a variety of features.
In total there are 14 features, including answer-based features, such as the number of \emph{words}, \emph{lines of code} and \emph{hyperlinks} to Web content the answer contains, the number of other answers the question already has, its readability score; as well as such as answerer's \emph{reputation}, how long the answerer has been registered (\emph{signup duration} in months) and as a percentile rank (\emph{signup percentile}), the number of answers they have previously written, time since the previous answer, the number of answers written by the answerer in his or her current session, and answer's position within the session.

\item \emph{Khan Academy}.
The online educational platform Khan Academy enables users learn a subject then practice what they learned through a sequence of questions on the subject.
We study performance during the practice stage by looking at whether users answered the questions correctly on their first attempt ($Y=1$) or not ($Y=0$). We study an anonymized sample of questions answered by adult Khan Academy users over a period from 07/2012 to 02/2014. For each question a user answers we have 19 features: 
as with Stack Exchange, these include answer-based, user-based, and other 
temporal features.

\item \emph{Duolingo}. The online language learning platform Duolingo is accessed through an app on a mobile device. Users are encouraged to use the app in short bursts during breaks and commutes. The data\footnote{\url{https://github.com/duolingo/halflife-regression}} was made available as part of a previous study~\cite{Settles2016}.
The data contains a 2-week sample 
(02/28/2013 through 03/12/2013) of ongoing activity of users learning a language. All users in this data started lessons before the beginning of the data collection period.
We focus on 
45K users who completed at least five lessons. The median number of lessons was 7, although some had as many as 639 lessons. Performance on a lesson is defined as $Y=1$ if the user got \emph{all} the words in the lesson correct; otherwise, it is $Y=0$. Features describing the user include how many lessons and sessions the user completed, how many perfect lessons the user had, the month and day of the lesson, etc.

\item \emph{Digg}.
The social news platform Digg allows users to post news stories, which their followers can like or ``digg.'' When a user diggs a story, that story is broadcast to his or her followers, 
a mechanism that allows for the diffusion of contents through the Digg social network. A further characteristic of Digg is its front page --- stories that are popular 
are promoted to the front page and thus become visible to \emph{every} Digg user. We study a dataset that tracks the diffusion of 3,500 popular Digg stories from their submissions by a single user to their eventual promotion to and residence on the Digg front page. We study information diffusion on Digg by examining whether or not ($Y \in \{0,1\}$) users ``digg'' (i.e., adopt) a story following exposure of the story from their friends, and thus share that information with their followers.  The features associated with adoption include user-based features, such as indegree and out degree (number of followers and followees of the users), node activity (how often the user posts), information received, (the rate at which the user receives information for all followees), dynamics-related features such as the the number of times the user was exposed to the story, story-related features such as its global popularity in the previous hour, and diurnal-features, including the hour of the day and day of the week. Through this data, we can study the factors that are important in the spread of information in this social system.

\item \emph{Twitter.}
On the online social network Twitter, users can post information, which is then broadcast to their followers, i.e., the other Twitter users that follow that user. This 
dataset tracks the spread of 65,000 unique URLs through the Twitter social network during 
one month in 2010. Similarly to \emph{Digg}, we can study social influence and information diffusion by examining whether ($Y=1$) or not ($Y=0$) a user posts a URL after being exposed to it when one of his or her friends posts. The features associated with each exposure event are the same as those for \emph{Digg}.

\end{itemize}

We compare the average predictive performance across 5 holdout sets of S3D to Lasso regression, elastic net regression, random forests, and linear SVM. We show that S3D can achieve competitive performance with the benchmark algorithms with a smaller set of features. Finally, we explore our tuned S3D models and demonstrate their utility to understanding 
human behaviors in \cref{sec:results_explore}. Relevant datasets and codes \footnote{\url{https://github.com/peterfennell/S3D/tree/paper-replication}} can be used to replicate the following results.

\subsection{Tuning hyperparameters}
 \begin{figure}
   \includegraphics[width=0.95\columnwidth]{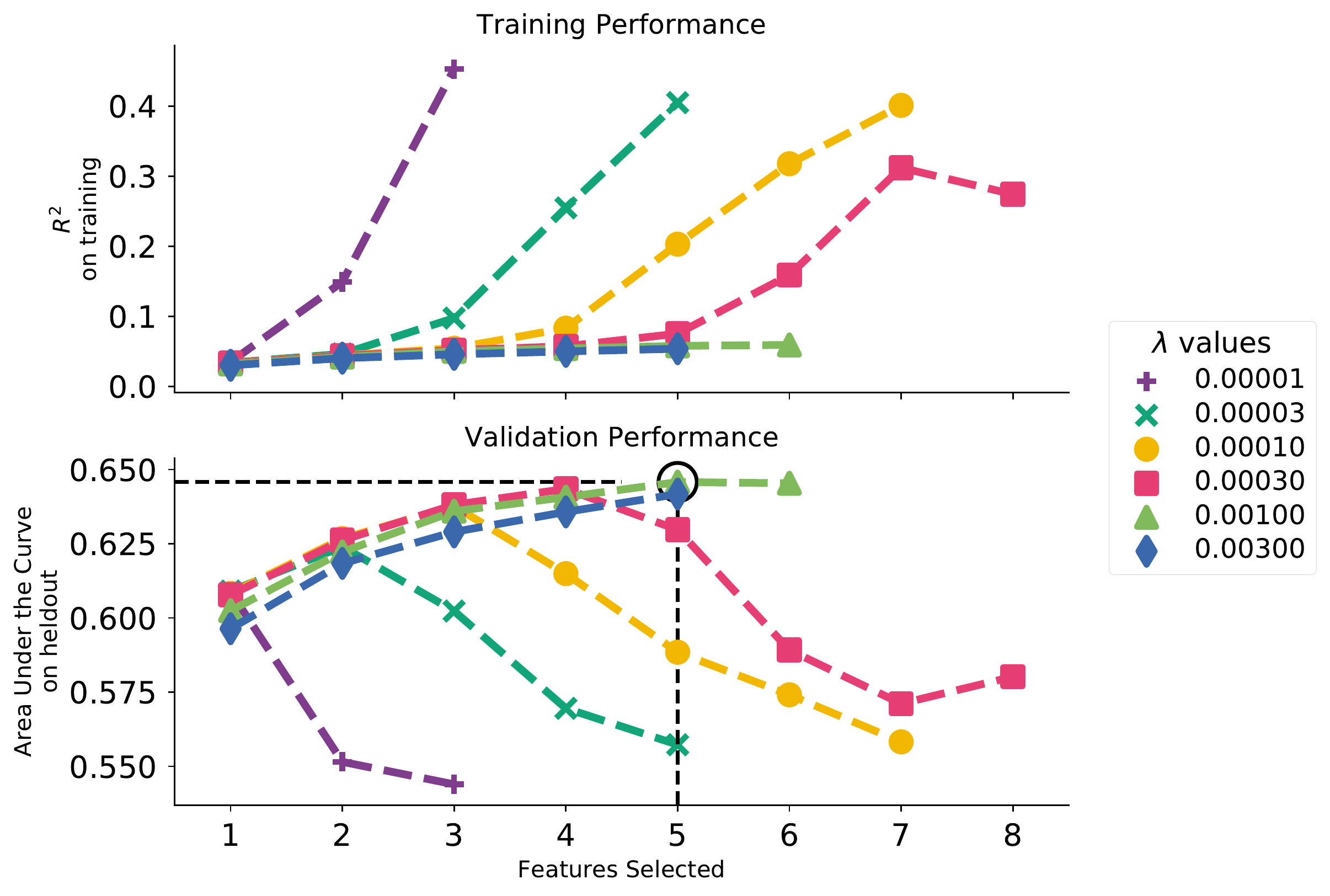}
 	\caption{Hyperparameter tuning on \emph{Stack Exchange} data. 
 Top: $R^2$ in the training set as a function of $\lambda$ and 
 $k$; Bottom: AUC on the held-out data. Note that for illustration, we show the trajectory for only one of five splits. Different ``best'' hyperparameters may exist for different splits.
 }
 	\label{fig:param_tuning}
 \end{figure}
An essential part of training a statistical model is hyperparameter tuning --- in the case of S3D, selecting the parameters $\lambda$ and 
$k$. This procedure is illustrated for Stack Exchange data in \cref{fig:param_tuning}, where we show the total $R^2$ at each step of the algorithm for various values of $\lambda$, as well as the 
AUCs at these steps computed on the held-out tuning data.
Overly small values of $\lambda$ perform quite poorly on the held-out data, as they produce very fine-grained partitions that overfit the data.
Larger values of $\lambda$ avoid being too fine-grained 
---the $R^2$ on the training set increases initially 
but diverges again as extra features selected in additional steps overfit the data (as shown through the decreasing performance on the held-out data; \cref{fig:param_tuning} bottom). Parameter 
$k$ (x-axis in \cref{fig:param_tuning}) 
controls the number of 
steps of S3D, thus picking the optimal model between underfitting and overfitting.
Supplementary file \emph{s3d\_hyperparameter\_df.csv}  \footnote{\url{https://raw.githubusercontent.com/peterfennell/S3D/paper-replication/replicate/s3d_hyperparameter_df.csv}} reports the best hyperparamters for all datasets, across the 4-fold inner CV processes.

\subsection{Prediction Performance}
\label{sec:results_benchmark_comparison}

 \begin{figure}[h]
 	\includegraphics[width=0.95\linewidth]{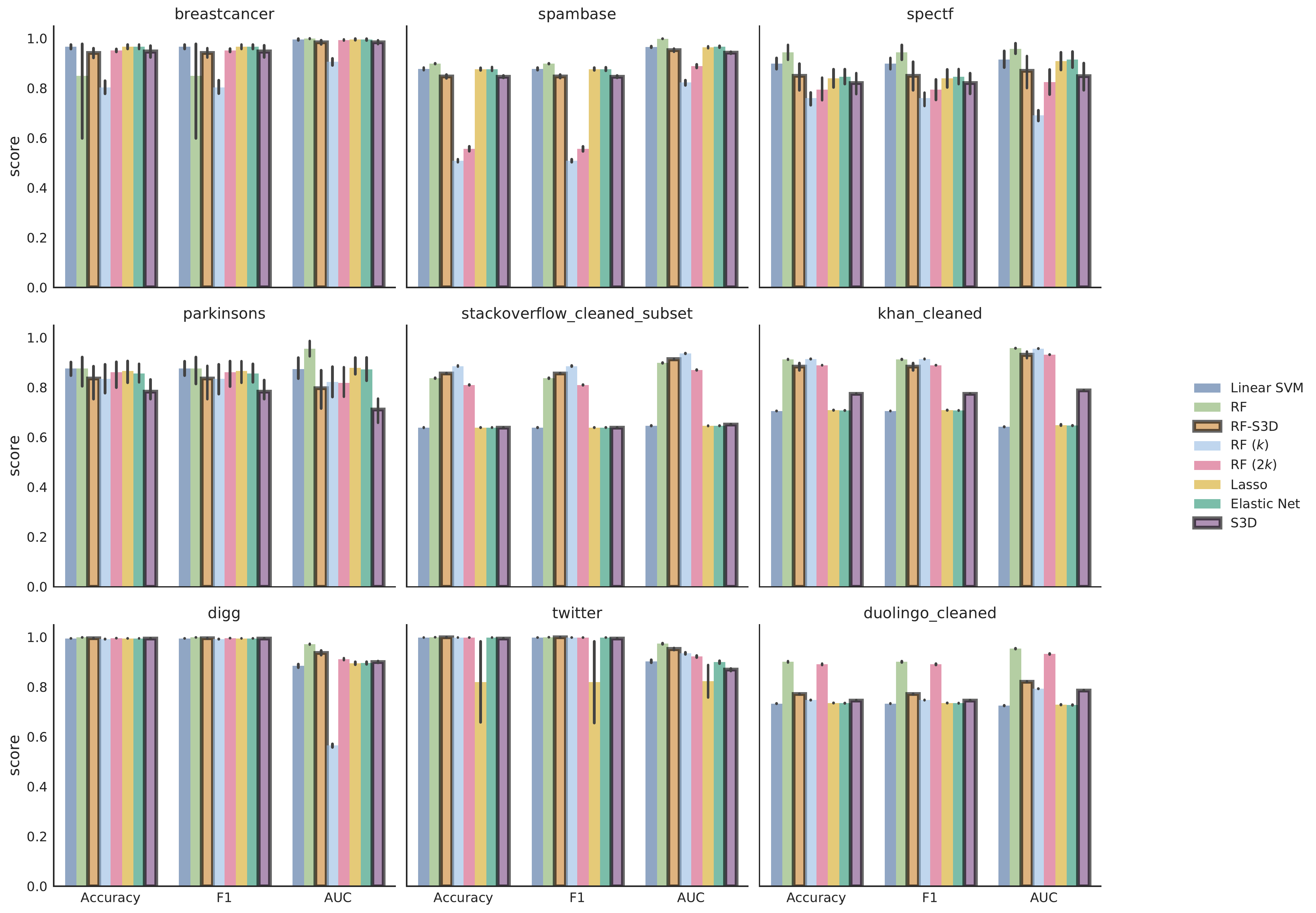}
 	\caption{Classification performance on 5-fold cross validation in 9 datasets. Error bars here indicate one standard deviation. A higher bar (greater value) means better performance.}
 	\label{fig:classification_comparison}
 \end{figure}

 \begin{figure}[h]
 	\includegraphics[width=0.95\linewidth]{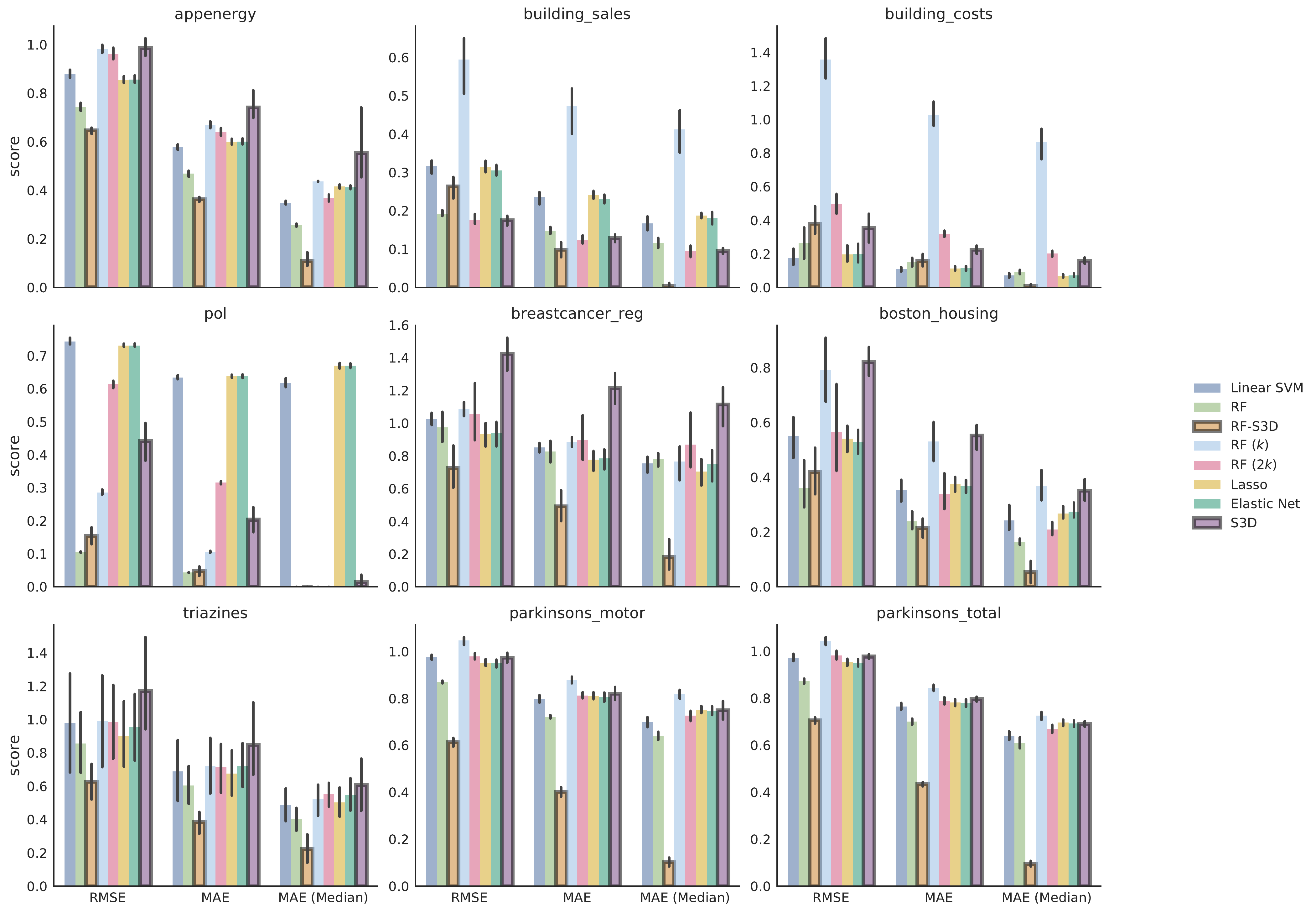}
 	\caption{Regression performance on 5-fold cross validation in 9 datasets. Error bars here indicate one standard deviation. A higher bar (greater value) means \emph{worse} performance.}
 	\label{fig:regression_comparison}
 \end{figure}

\cref{fig:classification_comparison,fig:regression_comparison} report performance on 
the outer CV for all datasets S3D, random forests, linear SVM, and logistic regressions (both Lasso and elastic net).
Overall S3D achieves predictive performance comparable to other state-of-the-art machine learning methods.

\begin{figure}
 	\includegraphics[width=0.95\linewidth]{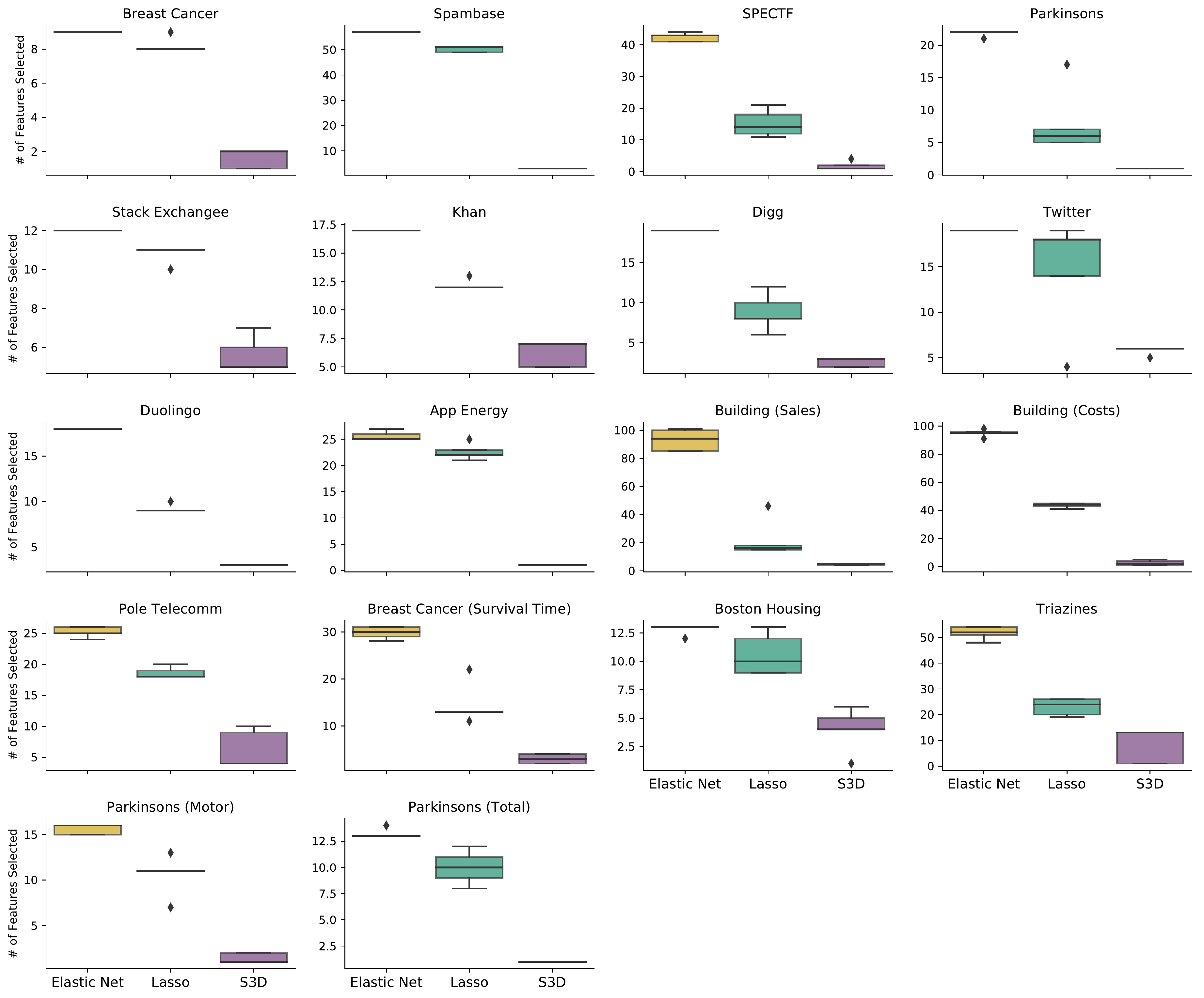}
 	\caption{Number of features selected by elastic net and lasso regression, as well as S3D, across outer CV for all datasets. The five points in each box correspond to the the number of features selected in each \emph{outer} fold.}
 	\label{fig:num_features}
 \end{figure}

In most cases, S3D's performance is similar to that of logistic regression and linear SVM. 
Its performance relative to random forest is especially remarkable considering the difference in  complexity of the models. S3D uses a subset of features and a simple $m$-dimensional hypercube to make predictions, in stark contrast to random forests, which use all features and learn many decision trees. In contrast, S3D selects a small set of features, producing more parsimonious models as compared to Lasso and elastic net regression (\cref{fig:num_features}).

S3D is especially useful as a feature selection method. Using just the few  features selected by S3D to train a random forest leads to highly competitive performance on the regression task for many datasets (RF-S3D bars in \cref{fig:regression_comparison}). Remarkably, its performance is often better than that of the random forest trained on the full feature set. This is likely because features selected by S3D are uncorrelated with each other; while correlations among features used by the random forest reduce performance.

\begin{figure}
 	\includegraphics[width=0.95\linewidth]{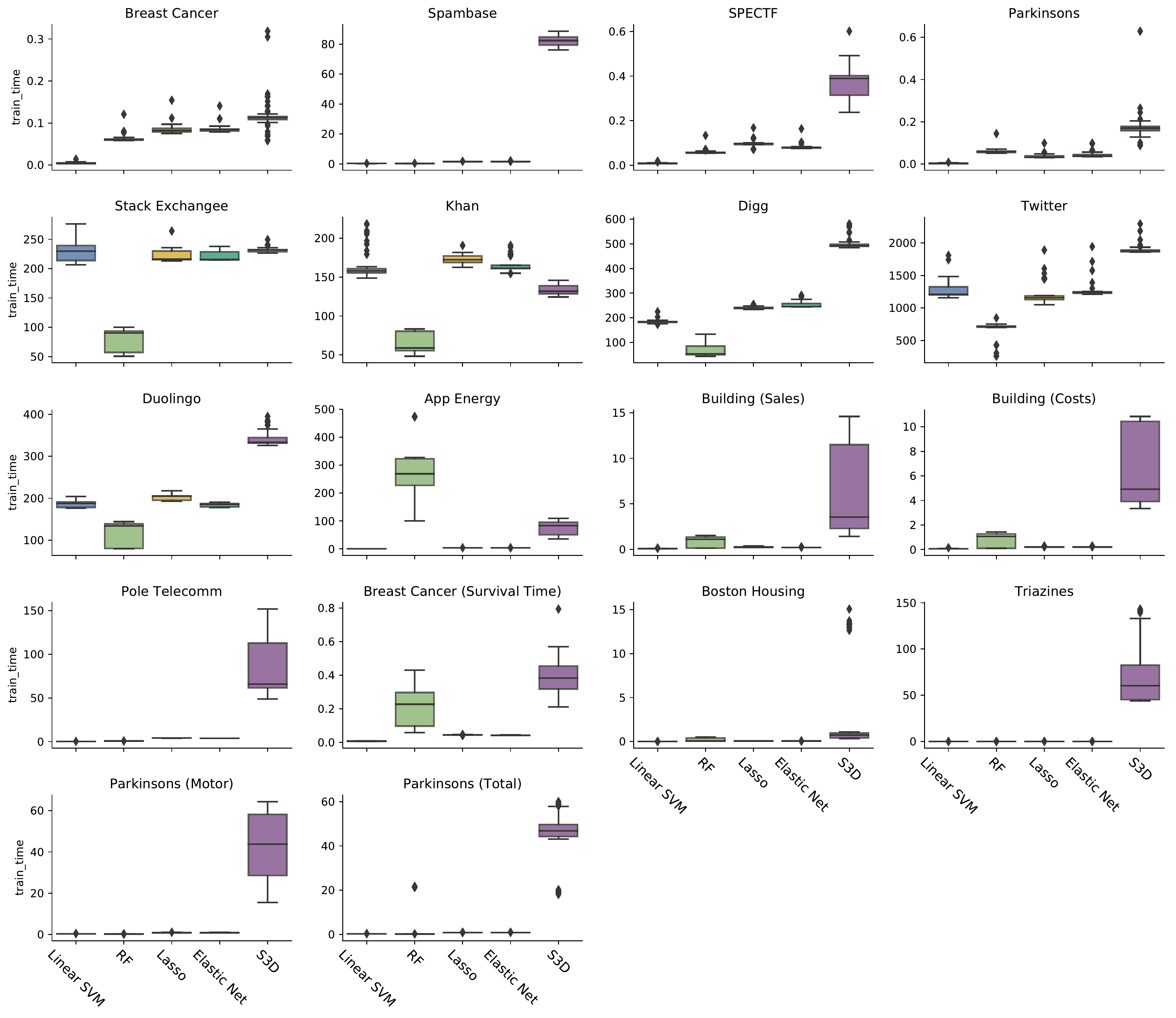}
 	\caption{Training time of all five algorithms.  Error bars indicate 25 and 75 percentiles, respectively.}
 	\label{fig:runtime}
 \end{figure}

Finally, we show that the runtime of S3D 
is competitive to the other four algorithms (\cref{fig:runtime}). For each dataset, all models were trained using the best parameters found in inner CV and full training sets over each split (recall that there are five splits) repeatedly ten times. Therefore, each box in \cref{fig:runtime} shows the distribution of training time over a total of 50 runs. {Note that the Python package Scikit Learn \cite{scikitlearn2011}} is used to implement logistic regression, random forest, and linear SVM, therefore producing superb runtime performance as it is highly optimized. We believe that the implementation of S3D can be further improved. For instance, the timing of S3D implementation includes reading the input file, whereas the other four have no such needs due to different setting.
Furthermore, \cref{fig:runtime} only reflects training time with one set of hyperparameters for each model. While random forests manifest outstanding efficiency, it is worth noting that the large amount of hyperparameters (in this study, we searched for four; there are at least four more) will inevitably lead to undesirably long hours of grid search. On other hand, S3D only needs two (i.e., $\lambda$ and $k$), which substantially reduce user effort in hyperparameter tuning.

\subsection{Analyzing human behavior with S3D}
In this section, we present a detailed description of applying S3D to understand human behaviors using \emph{Stack Exchange} data \footnote{See \url{https://github.com/peterfennell/S3D/blob/paper-replication/replicate/3-visualize-models.ipynb} for results on the other four human behavior datasets.}. We additionally included \emph{Digg} in \cref{sec:model_analysis} to demonstrate visualizations of the learned S3D model. Specifically, we used the best hyperparameters during cross validation: $\lambda=0.001$ and $k=5$ for \emph{Stack Exchange}; $\lambda=0.001$ and $k=3$ for \emph{Digg}. 
We note that, to explain the dataset, we applied S3D on \emph{all the data avaiable} using the best hyperparameters, which occurred most often during the cross validation processes.

\label{sec:results_explore}
\subsubsection{Feature Selection and Correlations}

For the large scale behavioral data, S3D selects a subset of features that collectively explain the largest amount of the variation in the outcome variable. In the meantime, it quantifies correlations between selected features to unselected ones. 
In the following, we describe selected features and examine the effects of the unselected ones.

 \begin{figure}
 	\includegraphics[width=0.98\textwidth]{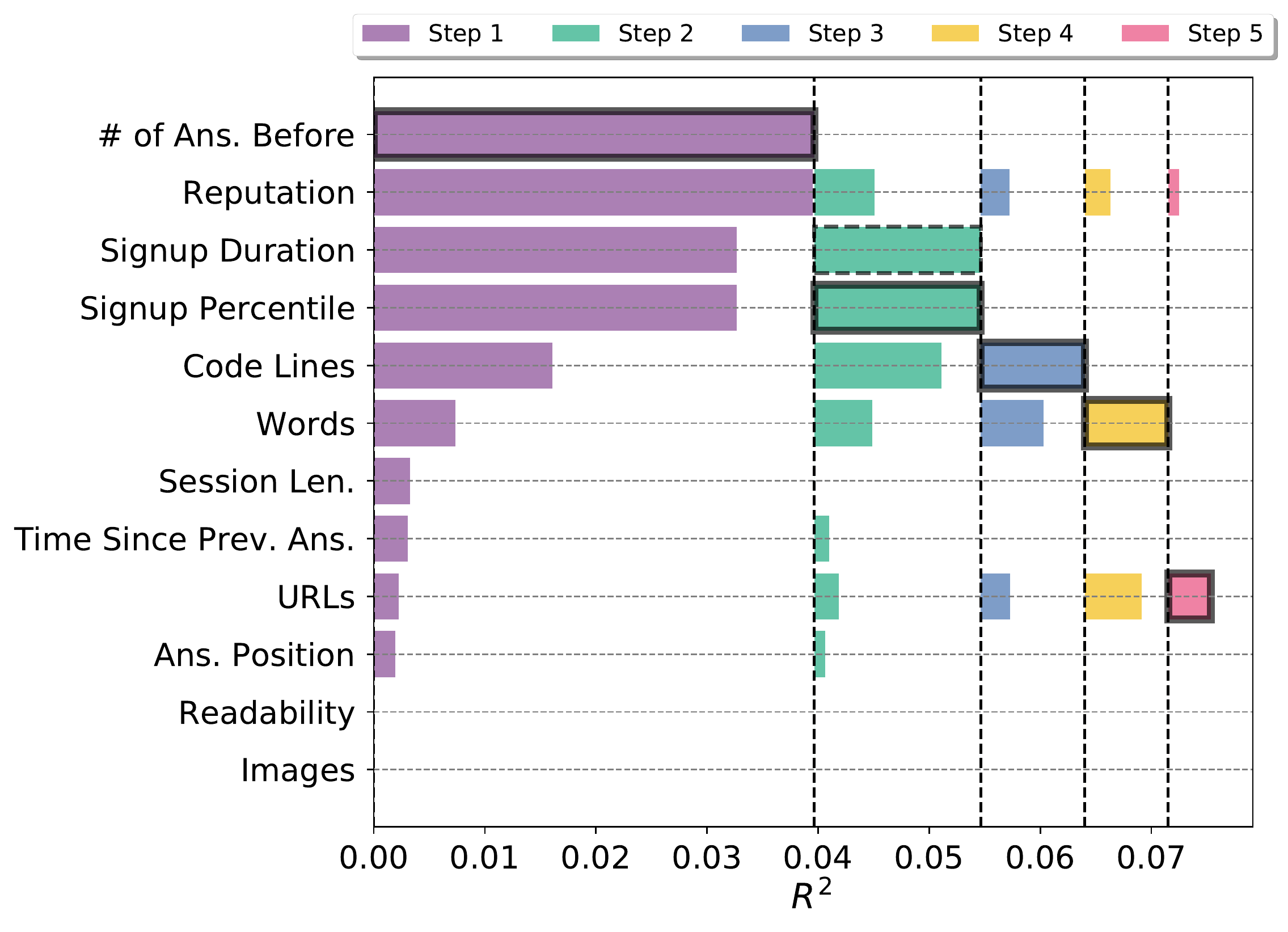}
 	\caption{The amount of variation in the Stack Exchange outcome variable \emph{performance} explained by each feature at each step of the S3D algorithm. The feature that explains most of the remaining variation at each step is highlighted here with a solid black rectangle surrounding its bar, whereas dashed rectangles indicate that the corresponding feature has the same amount of contribution (i.e., $\Delta R^2$) but not selected.
 	}
 	\label{fig:feat_importance}
 \end{figure}

\begin{figure}
 	\includegraphics[trim={2.5cm 1.7cm .3cm 1cm},clip,width=0.9\columnwidth]{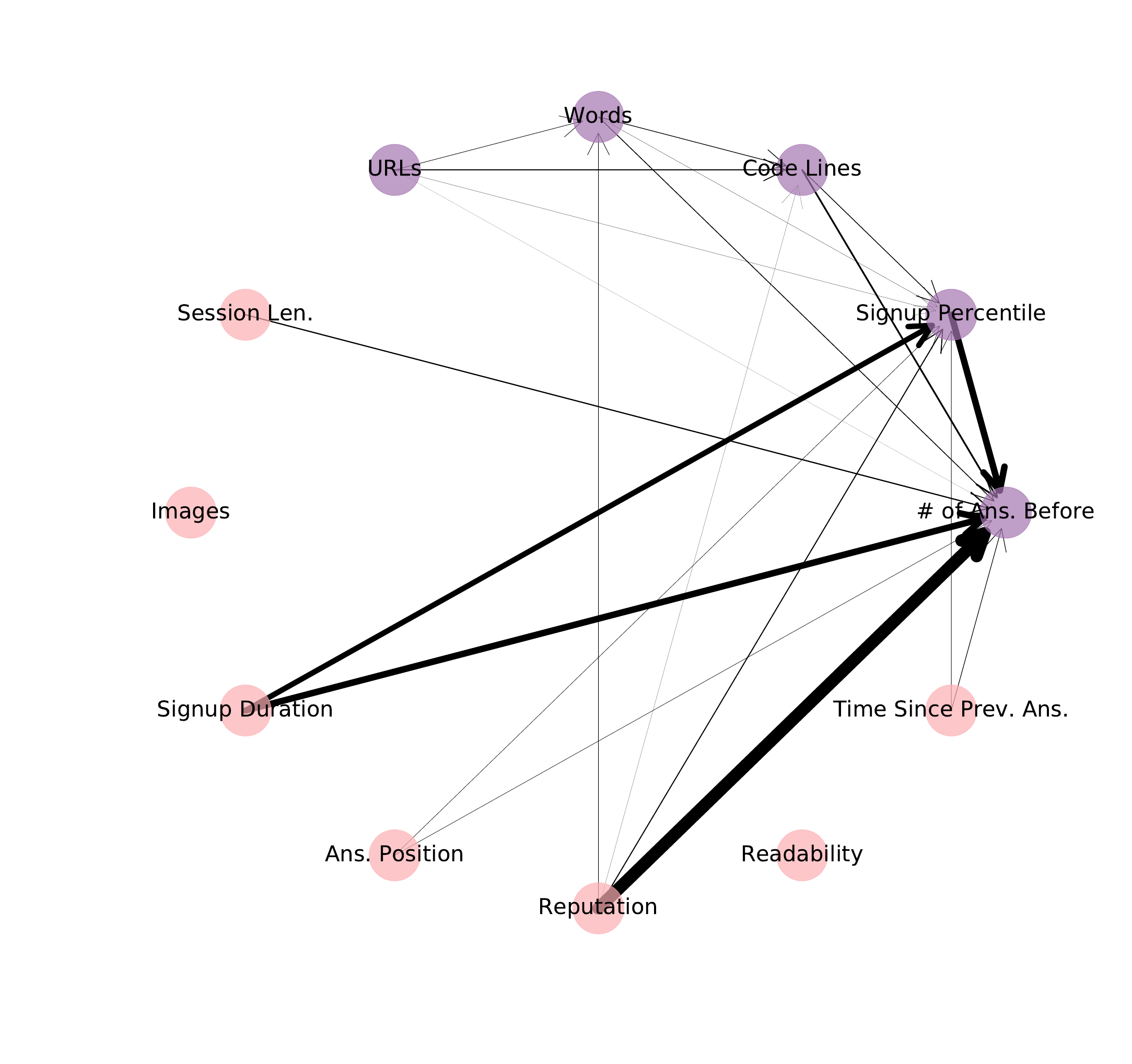}
 	\caption{The feature network for Stack Exchange, showing the variation in the outcome from the features that have not been selected (pink) through the selected features (purple). Edge width is proportional to weights described in \cref{sec:feat_selec_corr} --- the thicker a link between two features is, the more correlated they are.}
 	\label{fig:feat_net}
 \end{figure}

We give a detailed description of the features selected at each step (\cref{fig:feat_importance}) and the resulting feature network (\cref{fig:feat_net}). \cref{fig:feat_importance} visually ranks the features by showing the amount of variation explained by each feature at every step of the algorithm. The features selected at each step are outlined in black.
The first and most important feature selected is \emph{the number of answers provided before this question}. This feature, for one thing, indicates how active a user is in the community. For another, it implicitly reflects a user's capability. Interestingly, there is obvious dependencies between the number of previous answers and (\textit{1}) \textit{reputation}, (\textit{2}) \textit{signup duration/percentile}, and (\textit{3}) \textit{code lines}. Given the amount of previous answers in the model, the contribution of these features decrease dramatically.
The second feature S3D selects is \emph{signup percentile}, which measures answerers' ``age'' on Stack Exchange as a percentile rank. Intuitively, the longer a user stays in the system, the more likely they can accumulate their reputation and capability to produce a ``good'' answer.
It is noteworthy that \emph{signup duration} and \emph{percentile} share the exact amount of explained variation, which echoes the fact that the Spearman correlation between them is \textbf{\emph{1}}.
Following user tenure, the number of \textit{lines of codes} is selected as the third most important feature, followed by the \textit{number of words} and \textit{URLs}, which all, to some extent, manifest how informative an answer is. 
Note that the variation explained by the features \textit{number of words} and \textit{URLs} exceeds the variation explained by these features in the first step, leading to an interesting implication that there may exist an \emph{interaction} effect. In particular, given answers with the same \textit{number of code lines} and by answerers who signed up in similar time period and shared similar activeness, the \textit{number of words} and \textit{URLs} will contribute more to the final acceptance probability. The ability to identify moderation effects among variables, in fact, is a fascinating characteristic of S3D when analyzing heterogeneous behavioral data.

With $R^2=0.075$, the five selected features collectively explain the largest amount of variation in whether an answer is accepted by the asker as the best answer to his or her question. The unselected features have been made redundant by the selected features. Such redundancies can be represented as a directed and weighted network through the coefficients of \cref{eq:net_coeffs}, as shown in \cref{fig:feat_net}.
Specifically, links between selected features (purple) the unselected (pink) features show the variation in the outcome explained by the pink node can be explained by the purple node. The network visualizes the correlations and the significance of unselected features.
While some of the correlations are obvious, such as those between the \textit{number of answers}, user \textit{reputation}, and tenure length (i.e., \textit{signup duration}/\textit{percentile}), others are less evident. For example, there are links from \textit{reputation} to the number of 
\textit{words}, and \textit{code lines}, implying that reputable users may provide more detailed answers containing informative clues such as references to related webpages and sample codes. Although relatively weak, the link from answer position pointing towards the number of answers a user provided before and signup percentile alludes that senior users may tend to be more active and engaging in the community, therefore being early answerers to many questions.
The feature network, in this manner, not only lets us analyze which unselected features are explanatory of an outcome variable, but to which selected features they are correlated and are made redundant, providing a tool to suggest further exploration of correlations within the data.

In the Khan Academy dataset, S3D selects as important features: (\textit{1}) the \textit{time} it takes the user to solve the problem; (\textit{2}) the \textit{number of problems} that the user has solved on the first attempt without hints; (\textit{3}) \textit{time since previous problem}; (\textit{4}) the number of \textit{first five problems} solved correctly on the first attempt; (\textit{5}) \textit{index of the session} among all of that  user's sessions; (\textit{6}) \textit{index of the problem} within its session. It is noteworthy that the second and fourth features here are analogous to \textit{signup duration} and \textit{reputation} on Stack Exchange, as the number of problems that a user solves correctly on their first attempt is a combination of both skill and tenure.

For the Duolingo language learning platform, S3D picks similar skill-based features: (\textit{1}) the number of \textit{all lessons} completed perfectly; (\textit{2}) the number of \textit{prior lessons} completed; (\textit{3}) the number of \textit{distinct words} in the lesson. Similarly, the first feature here is equivalent to the second and fourth feature selected in Khan Academy, which quantifies both skill and tenure.

In the case of Digg social network, to explain whether a user will ``digg'' (or ``like'') a story recommended by friends, S3D selects  as important features:  (\textit{1}) user \textit{activity} (how many stories this user recommended); (\textit{2}) the amount of \textit{information received} by this user from the people she follows; and (\textit{3}) current popularity of this story 
The first two features describe how a user processes and receives information, while the third one reflects how ``viral'' a story is.

For Twitter, S3D selects (\textit{1}) the amount of \textit{information received} by this user; (\textit{2}) \textit{in-degree} (i.e., the number of followers and thus popularity) of friends; (\textit{3}) the number of \textit{times} this user has been exposed to this meme; (\textit{4}) user \textit{activity}; (\textit{5}) the age of this tweet; (\textit{6}) user's \textit{out-degree} (followees). S3D identifies the information received by the user as an important feature for both Digg and Twitter, which highlights important role that cognitive load plays in information spread online~\cite{Hodas2012}. On the other hand, the differences such as the lesser importance of user activity and greater importance of a user's friends in Twitter suggests interesting disparities in the manner of information diffusion on these two platforms.

\subsubsection{Model Analysis}
\label{sec:model_analysis}
	
One of the more novel contributions of S3D is its potential for data exploration. By iteratively selecting features and measuring the amount of outcome variation they explain (\cref{fig:feat_importance}), we can visualize the important dimensions of the model to fully understand both effects of important features and the corresponding predictions.
See \cref{supp-sec:visualization} for a more detailed step-by-step illustration.



\begin{figure}[!h]
    \includegraphics[width=.95\textwidth]{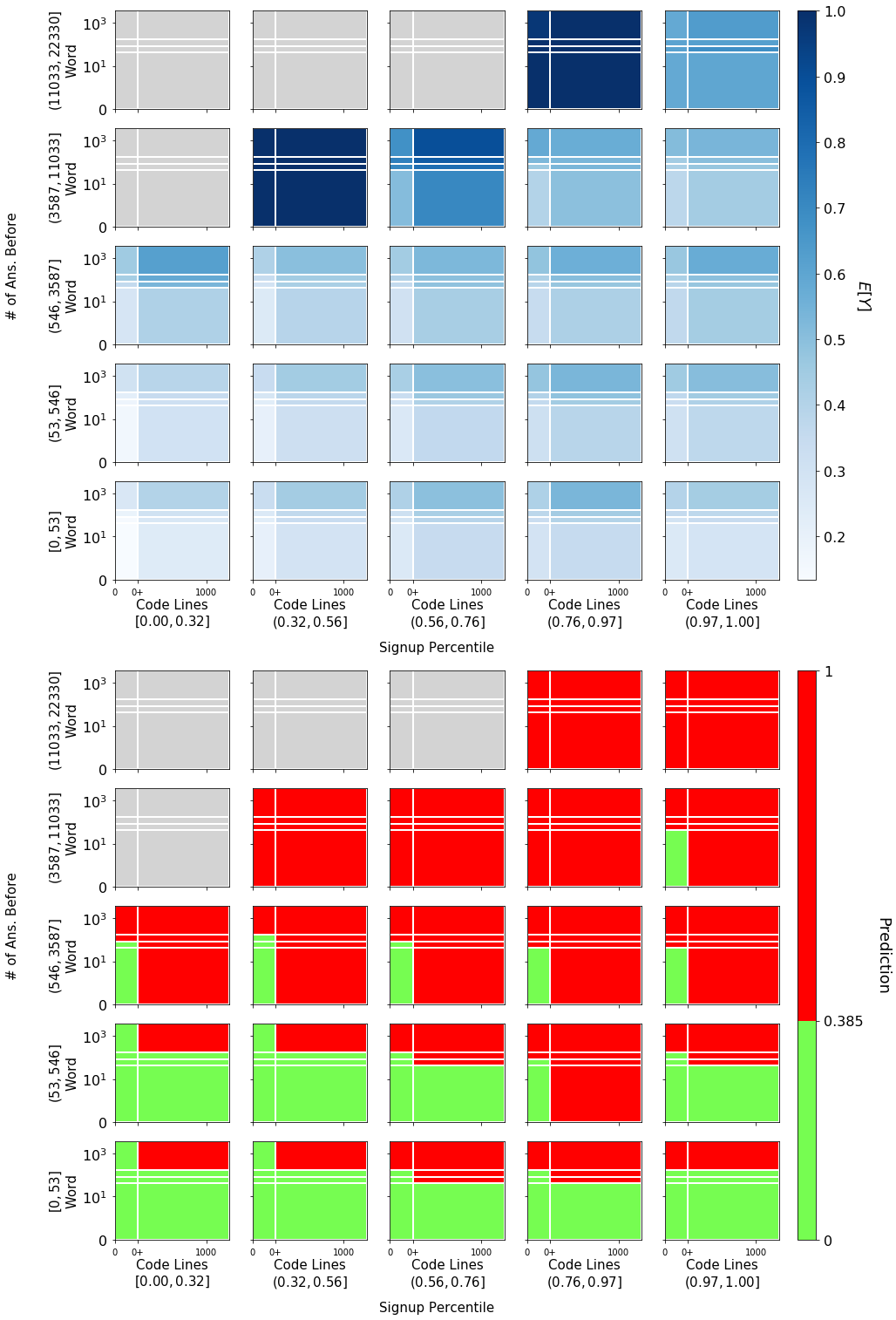}
  \caption{Visualization of the S3D model learned for Stack Exchange, showing the decomposition of feature space defined by the four most important features.
  These plots represent the partition of the 4-dimensional hypercube, and show how the acceptance probability (top) and corresponding predicted acceptance (bottom; red for label 1 and green for 0) of data points vary within within the space.
  \emph{Gray} areas have no observations.}
  \label{fig:interpretation}
\end{figure}


For Stack Exchange, S3D selects five important features. We visualized the model with the first four features in \cref{fig:interpretation},
that unfolds the $m=4$ hypercube learned by the model. It shows how the expectation (top plot) and the corresponding prediction (bottom plot) that the answer will be accepted as best answer, vary as a function of the four selected features. The prediction threshold selection is described in~\cref{sec:prediction}.
Each row of plots in~\cref{fig:interpretation} corresponds to a single bin of the first selected feature \emph{number of answers before}, while each column corresponds to bins of the second feature \emph{signup percentile rank}. Individual plots vary according to the third and fourth features \emph{code lines} and \emph{words}. It is quite evident that variation in 
the outcome (i.e., \cref{fig:interpretation} top plot) is greater between plots than within plots, a result of the fact that features are picked \emph{successively} to explain such variation. These plots show the collective effects of these four features: acceptance \emph{increases} with the user's experience (\emph{number of answers before} feature) and tenure (\emph{signup percentile rank}). Furthermore, longer answers with more \emph{words} and \emph{lines of code}  are more likely to be accepted as best answers.
Another interesting pattern emerges when the number of answers provided before is above 3,587: the acceptance rate rises when \textit{signup percentile} goes down. In other words, given a high level of user engagement in the community, newer users tend to produce answers that have higher chances of being accepted. On the other hand, more senior users tend to have a higher probability of having their answers accepted, when the number of previous answers is lower.

\begin{figure}
  \includegraphics[width=.96\columnwidth]{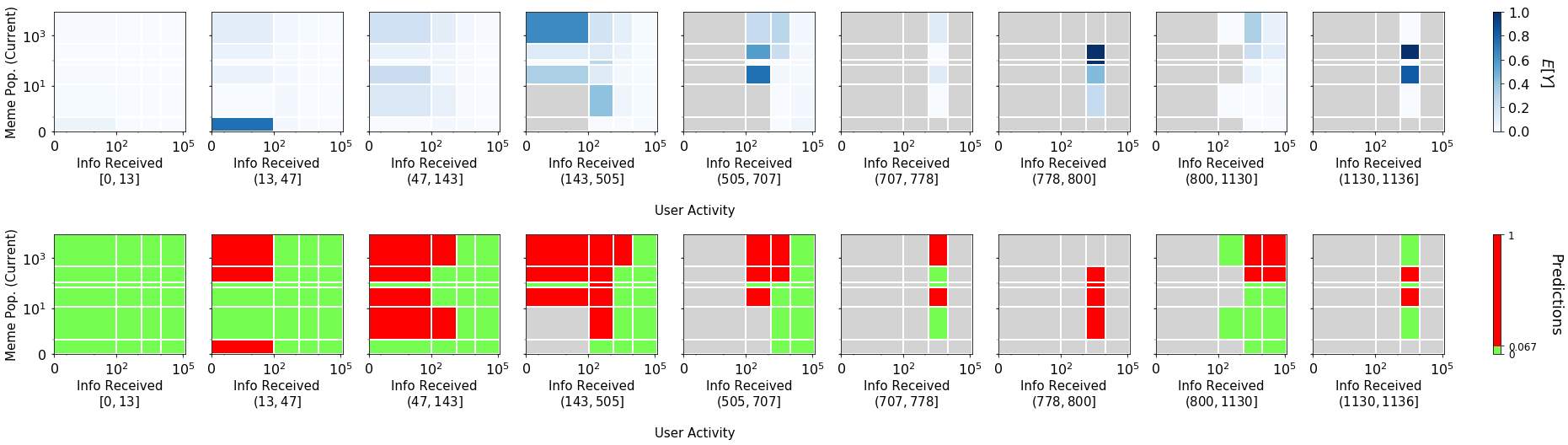}
  \caption{Visualization of the S3D model learned for Digg, showing the decomposition of feature space defined by \emph{three} most important features. Each plot shows adoption probability of a meme within each block in the 3D feature subspace. Each bottom plot shows predicted meme adoptions. \emph{Gray} areas have no observations.}
  \label{fig:digg}
\end{figure}

We also examine the S3D model learned for Digg to illustrate its effectiveness on highly unbalanced and heterogeneous data.
Here, S3D selects as important features \emph{user activity} (i.e., how often a user posts per day), \emph{information received} (the number of stories, or memes, a user's friends recommend), and
\emph{current popularity of the story}, i.e., how many users have recommended it.
The model, presented in \cref{fig:digg}, shows extreme heterogeneity in data with values of features and adoption probabilities varying widely, and notable here is S3D's ability to learn appropriate binnings of the features over many orders of magnitude.
Specifically, the figure shows that the probability of a user to adopt a story \emph{increases} when he/she is more active in the community (see also Figure S5), but \emph{decreases} as users receive more information from friends (see also Figure S6). Specifically, given  relatively low activity (e.g., adopting fewer than 505 stories),  those users seeing less information from friends are more likely to adopt a new story (corresponding to higher color intensity on the left hand side in each plot). The highly active users, on the other hand, also tend to receive more information---around 1,000 stories per day---from friends. However, they too are less likely to adopt a new story as they receive more information from friends. These two features---\textit{user activity} and \textit{information received}---represent the interplay between information processing and cognitive load. Our analysis highlights the extent of to which information overload, which occurs when users receive more information than they can effectively process, inhibits adoption and spread of memes online~\cite{Versteeg2011,Gomez-Rodriguez2014170,Hodas2012}.

The third feature \emph{current popularity} shows the impact of story popularity (i.e., virality or stickiness) on adoption. Our model shows that more popular stories are {more likely} to be adopted by individuals, as would be expected of viral memes. Striking is the absence of features related to the number or timing of exposures, either as selected features or in the feature network. The exposure effects may be quite subtle or even non-existent. The latter suggests that information on Digg spreads as a simple contagion where the probability of adopting a meme is independent of the number of exposures~\cite{Centola2007,Hodas2014}.

The large heterogeneity as a function of basic node features has important implications for the inference of social contagion, because heterogeneity and underlying confounders may distort analysis. A possible approach to such inference is to decompose the feature space, as in~\cref{fig:digg}, and statistically test the effect of multiple exposures in the resulting homogeneous blocks, an approach that would ensure that the most important factors that best explain the variation in the adoption of information have been conditioned on.


\section{Conclusion}
\label{sec:conclusion}

We introduced S3D, a statistical model with low complexity but strong predictive power and interpretability, that offers potential to greatly expand the scope of predictive models and machine learning. Our model gives a comprehensive description of the feature space, not only by selecting the most important features, but also by quantifying explained variation in the unselected features. This can be useful for practitioners, who are often concerned with the relationship between specific co-variates and an outcome variable.
The decomposed feature space learned by the model presents a powerful tool for exploratory data analysis,
and a medium for explaining model predictions. This is a positive step towards transparent algorithms that can be examined for bias, which presents a major stumbling block in the development and application of machine learning.

We have demonstrated the effectiveness of S3D on a variety of datasets, including benchmarks and real-world behavioral data, where it predicts outcomes in the held-out data at levels comparable to state-of-the-art machine learning algorithms. Its potential for interpreting complex behavioral data, however, goes beyond these alternate methods. S3D provides a ranking of features by their importance to explaining data, identifies correlated features, and can be readily used in visual data explorations. Our approach reveals the important factors in explaining human behavior, such as competition, skill, and answer complexity when analyzing performance on Stack Exchange or essential user attributes such as activity and information load in the social networks Digg and Twitter. Our method is self learning, involving minimal tuning of two hyperparameters, while also capable of systematically revealing the rich structure of complex data.

Aside from increasing our understanding of social systems, knowledge about what factors affect behavioral outcomes can also help us design of social platforms that improve human performance, including, for example, optimizing learning on educational platforms ~\cite{Rendle2012,Settles2016} or fairer judicial decisions~\cite{Kleinberg2017}.
The insights gained from the model can help design effective intervention strategies that change behaviors so as to improve individual and collective well-being.

Moving forward, there are two areas where S3D can make a considerable impact. The first is the mathematical modeling of human behavior,~\cite{Watts2002,Fernandez-Gracia2014,OSullivan2015}, where S3D can be used as a tool to extract model ingredients from data and help understand their functional form. The second is the use of S3D by practitioners to both explain predictions and analyze interventions based on these predictions. Transparency should be a key requirement for algorithms applied to sensitive areas such as predicting recidivism, and our work here shows that simple algorithms, such as S3D, can not only meet this requirement without sacrificing predictive accuracy. 
The development of machine learning tools should not be restricted to optimizing one single metric (predictive power), as other ingredients, such as interpretability, can 
improve how these methods are perceived by society.

\subsection*{Abbreviations}
  Structured Sum-of-Squares Decomposition (S3D). Total sum of squares (SST). Support Vector Machine (SVM). Cross validation (CV). Area under the curve (AUC). Random Forest (RF). Question-answering (Q\&A).


\begin{backmatter}
\section*{Declarations}

\subsection*{Availability of data and material}
The code and data required for replicating reported results are available here:
\url{https://github.com/peterfennell/S3D/tree/paper-replication}

\subsection*{Competing interests}
  The authors declare that they have no competing interests.

\subsection*{Funding}
This material is based upon work supported by the Defense Advanced Research Projects Agency (DARPA) and the Army Research Office (ARO) under Contracts No. W911NF-17-C-0094 and No. W911NF-18-C-0011, and in part by the James S. McDonnell Foundation. Any opinions, findings and conclusions or recommendations expressed in this material are those of the author(s) and do not necessarily reflect the views of DARPA and the ARO.

\subsection*{Author's contributions}
   PF designed and implemented the algorithm; ZZ carried out the validation studies; KL designed the algorithm. All authors participated in writing the paper.

\subsection*{Acknowledgements}
Authors are grateful to Raha Moraffa for her help setting up the cross validation framework and useful suggestions for improving performance.




\newcommand{\BMCxmlcomment}[1]{}

\BMCxmlcomment{

<refgrp>

<bibl id="B1">
  <title><p>Prediction and explanation in social systems</p></title>
  <aug>
    <au><snm>Hofman</snm><fnm>JM</fnm></au>
    <au><snm>Sharma</snm><fnm>A</fnm></au>
    <au><snm>Watts</snm><fnm>DJ</fnm></au>
  </aug>
  <source>Science</source>
  <publisher>American Association for the Advancement of Science</publisher>
  <pubdate>2017</pubdate>
  <volume>355</volume>
  <issue>6324</issue>
  <fpage>486</fpage>
  <lpage>-488</lpage>
  <url>http://dx.doi.org/10.1126/science.aal3856</url>
</bibl>

<bibl id="B2">
  <title><p>{A Trainable Spaced Repetition Model for Language
  Learning}</p></title>
  <aug>
    <au><snm>Settles</snm><fnm>B</fnm></au>
    <au><snm>Meeder</snm><fnm>B</fnm></au>
  </aug>
  <source>Proceedings of the 54th Annual Meeting of the Association for
  Computational Linguistics (Volume 1: Long Papers)</source>
  <publisher>Stroudsburg, PA, USA: Association for Computational
  Linguistics</publisher>
  <pubdate>2016</pubdate>
  <fpage>1848</fpage>
  <lpage>-1858</lpage>
</bibl>

<bibl id="B3">
  <title><p>Nudge: Improving Decisions About Health, Wealth, and
  Happiness</p></title>
  <aug>
    <au><snm>Thaler</snm><fnm>RH</fnm></au>
    <au><snm>Sunstein</snm><fnm>CR</fnm></au>
  </aug>
  <publisher>New York: Penguin Books</publisher>
  <edition>Rev and expanded</edition>
  <pubdate>2009</pubdate>
</bibl>

<bibl id="B4">
  <title><p>{Computational Social Science}</p></title>
  <aug>
    <au><snm>Lazer</snm><fnm>D</fnm></au>
    <au><snm>Pentland</snm><fnm>A</fnm></au>
    <au><snm>Adamic</snm><fnm>L</fnm></au>
    <au><snm>Aral</snm><fnm>S</fnm></au>
    <au><snm>Barabasi</snm><fnm>A. L.</fnm></au>
    <au><snm>Brewer</snm><fnm>D</fnm></au>
    <au><snm>Christakis</snm><fnm>N</fnm></au>
    <au><snm>Contractor</snm><fnm>N</fnm></au>
    <au><snm>Fowler</snm><fnm>J</fnm></au>
    <au><snm>Gutmann</snm><fnm>M</fnm></au>
    <au><snm>Jebara</snm><fnm>T</fnm></au>
    <au><snm>King</snm><fnm>G</fnm></au>
    <au><snm>Macy</snm><fnm>M</fnm></au>
    <au><snm>Roy</snm><fnm>D</fnm></au>
    <au><snm>{Van Alstyne}</snm><fnm>M</fnm></au>
  </aug>
  <source>Science</source>
  <publisher>American Association for the Advancement of Science</publisher>
  <pubdate>2009</pubdate>
  <volume>323</volume>
  <issue>5915</issue>
  <fpage>721</fpage>
  <lpage>-723</lpage>
</bibl>

<bibl id="B5">
  <title><p>{The Mythos of Model Interpretability}</p></title>
  <aug>
    <au><snm>Lipton</snm><fnm>ZC</fnm></au>
  </aug>
  <source>ACM Queue</source>
  <pubdate>2018</pubdate>
  <volume>16</volume>
  <issue>3</issue>
  <fpage>30</fpage>
</bibl>

<bibl id="B6">
  <title><p>{Prediction and explanation in social systems}</p></title>
  <aug>
    <au><snm>Hofman</snm><fnm>JM</fnm></au>
    <au><snm>Sharma</snm><fnm>A</fnm></au>
    <au><snm>Watts</snm><fnm>DJ</fnm></au>
  </aug>
  <source>Science</source>
  <pubdate>2017</pubdate>
  <volume>488</volume>
  <issue>February</issue>
  <fpage>486</fpage>
  <lpage>-488</lpage>
</bibl>

<bibl id="B7">
  <title><p>{Classification and regression trees}</p></title>
  <aug>
    <au><snm>Breiman</snm><fnm>L</fnm></au>
    <au><snm>Friedman</snm><fnm>JH</fnm></au>
    <au><snm>Olshen</snm><fnm>RA</fnm></au>
    <au><snm>Stone</snm><fnm>CJ</fnm></au>
  </aug>
  <publisher>New York: Wadsworth Publishing</publisher>
  <editor>Kimmel, John</editor>
  <edition>1</edition>
  <pubdate>1984</pubdate>
</bibl>

<bibl id="B8">
  <title><p>{On Simpson's Paradox and the Sure-Thing Principle}</p></title>
  <aug>
    <au><snm>Blyth</snm><fnm>CR</fnm></au>
  </aug>
  <source>Journal of the American Statistical Association</source>
  <pubdate>1972</pubdate>
  <volume>67</volume>
  <issue>338</issue>
  <fpage>364</fpage>
  <lpage>-366</lpage>
</bibl>

<bibl id="B9">
  <title><p>{Can you Trust the Trend: Discovering Simpson's Paradoxes in Social
  Data}</p></title>
  <aug>
    <au><snm>Alipourfard</snm><fnm>N</fnm></au>
    <au><snm>Fennell</snm><fnm>PG</fnm></au>
    <au><snm>Lerman</snm><fnm>K</fnm></au>
  </aug>
  <source>Proceedings of the Eleventh ACM International Conference on Web
  Search and Data Mining - WSDM '18</source>
  <publisher>New York, New York, USA: ACM Press</publisher>
  <pubdate>2018</pubdate>
  <fpage>19</fpage>
  <lpage>-27</lpage>
</bibl>

<bibl id="B10">
  <title><p>{The Elements of Statistical Learning}</p></title>
  <aug>
    <au><snm>Hastie</snm><fnm>T</fnm></au>
    <au><snm>Tibshirani</snm><fnm>R</fnm></au>
    <au><snm>Friedman</snm><fnm>J</fnm></au>
  </aug>
  <publisher>New York, NY: Springer New York</publisher>
  <edition>2</edition>
  <series><title><p>Springer Series in Statistics</p></title></series>
  <pubdate>2009</pubdate>
</bibl>

<bibl id="B11">
  <title><p>{Regularization and variable selection via the elastic
  net}</p></title>
  <aug>
    <au><snm>Zou</snm><fnm>H</fnm></au>
    <au><snm>Hastie</snm><fnm>T</fnm></au>
  </aug>
  <source>Journal of the Royal Statistical Society: Series B (Statistical
  Methodology)</source>
  <pubdate>2005</pubdate>
  <volume>67</volume>
  <issue>2</issue>
  <fpage>301</fpage>
  <lpage>-320</lpage>
</bibl>

<bibl id="B12">
  <title><p>{BART: Bayesian additive regression trees}</p></title>
  <aug>
    <au><snm>Chipman</snm><fnm>HA</fnm></au>
    <au><snm>George</snm><fnm>EI</fnm></au>
    <au><snm>McCulloch</snm><fnm>RE</fnm></au>
  </aug>
  <source>The Annals of Applied Statistics</source>
  <publisher>Institute of Mathematical Statistics</publisher>
  <pubdate>2010</pubdate>
  <volume>4</volume>
  <issue>1</issue>
  <fpage>266</fpage>
  <lpage>-298</lpage>
</bibl>

<bibl id="B13">
  <title><p>{Multivariate Adaptive Regression Splines}</p></title>
  <aug>
    <au><snm>Friedman</snm><fnm>J</fnm></au>
  </aug>
  <source>The Annals of Statistics</source>
  <publisher>Institute of Mathematical Statistics</publisher>
  <pubdate>1991</pubdate>
  <volume>19</volume>
  <issue>1</issue>
  <fpage>1</fpage>
  <lpage>-67</lpage>
</bibl>

<bibl id="B14">
  <title><p>{Random forests}</p></title>
  <aug>
    <au><snm>Breiman</snm><fnm>L</fnm></au>
  </aug>
  <source>Machine Learning</source>
  <publisher>Kluwer Academic Publishers</publisher>
  <pubdate>2001</pubdate>
  <volume>45</volume>
  <issue>1</issue>
  <fpage>5</fpage>
  <lpage>-32</lpage>
</bibl>

<bibl id="B15">
  <title><p>{A training algorithm for optimal margin classifiers}</p></title>
  <aug>
    <au><snm>Boser</snm><fnm>BE</fnm></au>
    <au><snm>Guyon</snm><fnm>IM</fnm></au>
    <au><snm>Vapnik</snm><fnm>VN</fnm></au>
  </aug>
  <source>Proceedings of the fifth annual workshop on Computational learning
  theory - COLT '92</source>
  <publisher>New York, New York, USA: ACM Press</publisher>
  <pubdate>1992</pubdate>
  <fpage>144</fpage>
  <lpage>-152</lpage>
</bibl>

<bibl id="B16">
  <title><p>{Do we Need Hundreds of Classifiers to Solve Real World
  Classification Problems?}</p></title>
  <aug>
    <au><snm>Fern{\'{a}}ndez Delgado</snm><fnm>M</fnm></au>
    <au><snm>Cernadas</snm><fnm>E</fnm></au>
    <au><snm>Barro</snm><fnm>S</fnm></au>
    <au><snm>Amorim</snm><fnm>D</fnm></au>
    <au><snm>Fern{\'{a}}ndez Delgado</snm><fnm>A</fnm></au>
  </aug>
  <source>Journal of Machine Learning Research</source>
  <pubdate>2014</pubdate>
  <volume>15</volume>
  <fpage>3133</fpage>
  <lpage>-3181</lpage>
</bibl>

<bibl id="B17">
  <title><p>{Training and Testing Low-degree Polynomial Data Mappings via
  Linear SVM}</p></title>
  <aug>
    <au><snm>Chang</snm><fnm>YW</fnm></au>
    <au><snm>Hsieh</snm><fnm>CJ</fnm></au>
    <au><snm>Chang</snm><fnm>KW</fnm></au>
    <au><snm>Ringgaard</snm><fnm>M</fnm></au>
    <au><snm>Lin</snm><fnm>CJ</fnm></au>
  </aug>
  <source>Journal of Machine Learning Research</source>
  <pubdate>2010</pubdate>
  <volume>11</volume>
  <issue>Apr</issue>
  <fpage>1471</fpage>
  <lpage>-1490</lpage>
</bibl>

<bibl id="B18">
  <title><p>{{\{}UCI{\}} Machine Learning Repository}</p></title>
  <aug>
    <au><snm>Dheeru</snm><fnm>D</fnm></au>
    <au><snm>{Karra Taniskidou}</snm><fnm>E</fnm></au>
  </aug>
  <pubdate>2017</pubdate>
  <url>http://archive.ics.uci.edu/ml</url>
</bibl>

<bibl id="B19">
  <title><p>{Nuclear feature extraction for breast tumor diagnosis}</p></title>
  <aug>
    <au><snm>Street</snm><fnm>W.N.</fnm></au>
    <au><snm>Wolberg</snm><fnm>W.H.</fnm></au>
    <au><snm>Mangasarian</snm><fnm>O.L.</fnm></au>
  </aug>
  <source>ISandT/SPIE International Symposium on Electronic Imaging: Science
  and Technology</source>
  <pubdate>1993</pubdate>
  <volume>1905</volume>
  <fpage>861</fpage>
  <lpage>-870</lpage>
</bibl>

<bibl id="B20">
  <title><p>{Knowledge discovery approach to automated cardiac SPECT
  diagnosis}</p></title>
  <aug>
    <au><snm>Kurgan</snm><fnm>LA</fnm></au>
    <au><snm>Cios</snm><fnm>KJ</fnm></au>
    <au><snm>Tadeusiewicz</snm><fnm>R</fnm></au>
    <au><snm>Ogiela</snm><fnm>M</fnm></au>
    <au><snm>Goodenday</snm><fnm>LS</fnm></au>
  </aug>
  <source>Artificial Intelligence in Medicine</source>
  <publisher>Elsevier</publisher>
  <pubdate>2001</pubdate>
  <volume>23</volume>
  <issue>2</issue>
  <fpage>149</fpage>
  <lpage>-169</lpage>
</bibl>

<bibl id="B21">
  <title><p>{Exploiting nonlinear recurrence and fractal scaling properties for
  voice disorder detection}</p></title>
  <aug>
    <au><snm>Little</snm><fnm>MA</fnm></au>
    <au><snm>McSharry</snm><fnm>PE</fnm></au>
    <au><snm>Roberts</snm><fnm>SJ</fnm></au>
    <au><snm>Costello</snm><fnm>DA</fnm></au>
    <au><snm>Moroz</snm><fnm>IM</fnm></au>
  </aug>
  <source>BioMedical Engineering Online</source>
  <publisher>BioMed Central</publisher>
  <pubdate>2007</pubdate>
  <volume>6</volume>
  <issue>1</issue>
  <fpage>23</fpage>
</bibl>

<bibl id="B22">
  <title><p>{Data driven prediction models of energy use of appliances in a
  low-energy house}</p></title>
  <aug>
    <au><snm>Candanedo</snm><fnm>LM</fnm></au>
    <au><snm>Feldheim</snm><fnm>V</fnm></au>
    <au><snm>Deramaix</snm><fnm>D</fnm></au>
  </aug>
  <source>Energy and Buildings</source>
  <publisher>Elsevier</publisher>
  <pubdate>2017</pubdate>
  <volume>140</volume>
  <fpage>81</fpage>
  <lpage>-97</lpage>
</bibl>

<bibl id="B23">
  <title><p>{A Novel Machine Learning Model for Estimation of Sale Prices of
  Real Estate Units}</p></title>
  <aug>
    <au><snm>Rafiei</snm><fnm>MH</fnm></au>
    <au><snm>Adeli</snm><fnm>H</fnm></au>
  </aug>
  <source>Journal of Construction Engineering and Management</source>
  <pubdate>2016</pubdate>
  <volume>142</volume>
  <issue>2</issue>
  <fpage>04015066</fpage>
</bibl>

<bibl id="B24">
  <title><p>{Rule-based Machine Learning Methods for Functional
  Prediction}</p></title>
  <aug>
    <au><snm>Weiss</snm><fnm>S M</fnm></au>
    <au><snm>Indurkhya</snm><fnm>N</fnm></au>
  </aug>
  <source>Journal of Artificial Intelligence Research</source>
  <pubdate>1995</pubdate>
  <volume>3</volume>
  <issue>1995</issue>
  <fpage>383</fpage>
  <lpage>-403</lpage>
</bibl>

<bibl id="B25">
  <title><p>{Hedonic housing prices and the demand for clean air}</p></title>
  <aug>
    <au><snm>Harrison</snm><fnm>D</fnm></au>
    <au><snm>Rubinfeld</snm><fnm>DL</fnm></au>
  </aug>
  <source>Journal of Environmental Economics and Management</source>
  <publisher>Academic Press</publisher>
  <pubdate>1978</pubdate>
  <volume>5</volume>
  <issue>1</issue>
  <fpage>81</fpage>
  <lpage>-102</lpage>
</bibl>

<bibl id="B26">
  <title><p>{Comparison of artificial intelligence methods for modeling
  pharmaceutical QSARS}</p></title>
  <aug>
    <au><snm>King</snm><fnm>RD</fnm></au>
    <au><snm>Hirst</snm><fnm>JD</fnm></au>
    <au><snm>Sternberg</snm><fnm>MJE</fnm></au>
  </aug>
  <source>Applied Artificial Intelligence</source>
  <publisher>Taylor {\&} Francis Group</publisher>
  <pubdate>1995</pubdate>
  <volume>9</volume>
  <issue>2</issue>
  <fpage>213</fpage>
  <lpage>-233</lpage>
</bibl>

<bibl id="B27">
  <title><p>{Suitability of Dysphonia Measurements for Telemonitoring of
  Parkinson's Disease}</p></title>
  <aug>
    <au><snm>Little</snm><fnm>M.A.</fnm></au>
    <au><snm>McSharry</snm><fnm>P.E.</fnm></au>
    <au><snm>Hunter</snm><fnm>E.J.</fnm></au>
    <au><snm>Spielman</snm><fnm>J</fnm></au>
    <au><snm>Ramig</snm><fnm>L.O.</fnm></au>
  </aug>
  <source>IEEE Transactions on Biomedical Engineering</source>
  <pubdate>2009</pubdate>
  <volume>56</volume>
  <issue>4</issue>
  <fpage>1015</fpage>
  <lpage>-1022</lpage>
</bibl>

<bibl id="B28">
  <title><p>{Scikit-learn: Machine Learning in Python}</p></title>
  <aug>
    <au><snm>Pedregosa</snm><fnm>F</fnm></au>
    <au><snm>Varoquaux</snm><fnm>G</fnm></au>
    <au><snm>Gramfort</snm><fnm>A</fnm></au>
    <au><snm>Michel</snm><fnm>V</fnm></au>
    <au><snm>Thirion</snm><fnm>B</fnm></au>
    <au><snm>Grisel</snm><fnm>O</fnm></au>
    <au><snm>Blondel</snm><fnm>M</fnm></au>
    <au><snm>Prettenhofer</snm><fnm>P</fnm></au>
    <au><snm>Weiss</snm><fnm>R</fnm></au>
    <au><snm>Dubourg</snm><fnm>V</fnm></au>
    <au><snm>Vanderplas</snm><fnm>J</fnm></au>
    <au><snm>Passos</snm><fnm>A</fnm></au>
    <au><snm>Cournapeau</snm><fnm>D</fnm></au>
    <au><snm>Brucher</snm><fnm>M</fnm></au>
    <au><snm>Perrot</snm><fnm>M</fnm></au>
    <au><snm>Duchesnay</snm><fnm>{\'{E}}</fnm></au>
  </aug>
  <source>Journal of Machine Learning Research</source>
  <pubdate>2011</pubdate>
  <volume>12</volume>
  <issue>Oct</issue>
  <fpage>2825</fpage>
  <lpage>-2830</lpage>
  <url>http://jmlr.csail.mit.edu/papers/v12/pedregosa11a.html</url>
</bibl>

<bibl id="B29">
  <title><p>{How visibility and divided attention constrain social
  contagion}</p></title>
  <aug>
    <au><snm>Hodas</snm><fnm>NO</fnm></au>
    <au><snm>Lerman</snm><fnm>K</fnm></au>
  </aug>
  <source>Proceedings - 2012 ASE/IEEE International Conference on Privacy,
  Security, Risk and Trust and 2012 ASE/IEEE International Conference on Social
  Computing, SocialCom/PASSAT 2012</source>
  <publisher>IEEE</publisher>
  <pubdate>2012</pubdate>
  <fpage>249</fpage>
  <lpage>-257</lpage>
</bibl>

<bibl id="B30">
  <title><p>What stops social epidemics?</p></title>
  <aug>
    <au><cnm>{Greg {Ver Steeg}}</cnm></au>
    <au><snm>Ghosh</snm><fnm>R</fnm></au>
    <au><snm>Lerman</snm><fnm>K</fnm></au>
  </aug>
  <source>Proceedings of 5th International Conference on Weblogs and Social
  Media</source>
  <pubdate>2011</pubdate>
</bibl>

<bibl id="B31">
  <title><p>{Quantifying information overload in social media and its impact on
  social contagions}</p></title>
  <aug>
    <au><snm>Gomez Rodriguez</snm><fnm>M</fnm></au>
    <au><snm>Gummadi</snm><fnm>K P</fnm></au>
    <au><snm>Sch{\"{o}}lkopf</snm><fnm>B</fnm></au>
  </aug>
  <source>Proceedings of the 8th International Conference on Weblogs and Social
  Media, ICWSM 2014</source>
  <pubdate>2014</pubdate>
  <fpage>170</fpage>
  <lpage>-179</lpage>
</bibl>

<bibl id="B32">
  <title><p>{Cascade dynamics of complex propagation}</p></title>
  <aug>
    <au><snm>Centola</snm><fnm>D</fnm></au>
    <au><snm>Egu{\'{i}}luz</snm><fnm>VM</fnm></au>
    <au><snm>Macy</snm><fnm>MW</fnm></au>
  </aug>
  <source>Physica A: Statistical Mechanics and its Applications</source>
  <publisher>North-Holland</publisher>
  <pubdate>2007</pubdate>
  <volume>374</volume>
  <issue>1</issue>
  <fpage>449</fpage>
  <lpage>-456</lpage>
</bibl>

<bibl id="B33">
  <title><p>{The simple rules of social contagion}</p></title>
  <aug>
    <au><snm>Hodas</snm><fnm>NO</fnm></au>
    <au><snm>Lerman</snm><fnm>K</fnm></au>
  </aug>
  <source>Scientific Reports</source>
  <publisher>Nature Publishing Group</publisher>
  <pubdate>2014</pubdate>
  <volume>4</volume>
  <issue>1</issue>
  <fpage>4343</fpage>
</bibl>

<bibl id="B34">
  <title><p>{Factorization Machines with libFM}</p></title>
  <aug>
    <au><snm>Rendle</snm><fnm>S</fnm></au>
  </aug>
  <source>ACM Transactions on Intelligent Systems and Technology</source>
  <publisher>IEEE</publisher>
  <pubdate>2012</pubdate>
  <volume>3</volume>
  <issue>3</issue>
  <fpage>1</fpage>
  <lpage>-22</lpage>
</bibl>

<bibl id="B35">
  <title><p>{Human Decisions and Machine Predictions}</p></title>
  <aug>
    <au><snm>Kleinberg</snm><fnm>J</fnm></au>
    <au><snm>Lakkaraju</snm><fnm>H</fnm></au>
    <au><snm>Leskovec</snm><fnm>J</fnm></au>
    <au><snm>Ludwig</snm><fnm>J</fnm></au>
    <au><snm>Mullainathan</snm><fnm>S</fnm></au>
  </aug>
  <source>The Quarterly Journal of Economics</source>
  <pubdate>2017</pubdate>
  <volume>133</volume>
  <issue>1</issue>
  <fpage>237</fpage>
  <lpage>-293</lpage>
</bibl>

<bibl id="B36">
  <title><p>{A simple model of global cascades on random networks}</p></title>
  <aug>
    <au><snm>Watts</snm><fnm>D. J.</fnm></au>
  </aug>
  <source>Proceedings of the National Academy of Sciences</source>
  <publisher>National Academy of Sciences</publisher>
  <pubdate>2002</pubdate>
  <volume>99</volume>
  <issue>9</issue>
  <fpage>5766</fpage>
  <lpage>-5771</lpage>
</bibl>

<bibl id="B37">
  <title><p>{Is the Voter Model a Model for Voters?}</p></title>
  <aug>
    <au><snm>Fern{\'{a}}ndez Gracia</snm><fnm>J</fnm></au>
    <au><snm>Suchecki</snm><fnm>K</fnm></au>
    <au><snm>Ramasco</snm><fnm>JJ</fnm></au>
    <au><snm>{San Miguel}</snm><fnm>M</fnm></au>
    <au><snm>Egu{\'{i}}luz</snm><fnm>VM</fnm></au>
  </aug>
  <source>Physical Review Letters</source>
  <publisher>American Physical Society</publisher>
  <pubdate>2014</pubdate>
  <volume>112</volume>
  <issue>15</issue>
  <fpage>158701</fpage>
</bibl>

<bibl id="B38">
  <title><p>{Mathematical modeling of complex contagion on clustered
  networks}</p></title>
  <aug>
    <au><snm>O'Sullivan</snm><fnm>DJP</fnm></au>
    <au><snm>O'Keeffe</snm><fnm>GJ</fnm></au>
    <au><snm>Fennell</snm><fnm>PG</fnm></au>
    <au><snm>Gleeson</snm><fnm>JP</fnm></au>
  </aug>
  <source>Frontiers in Physics</source>
  <publisher>Frontiers</publisher>
  <pubdate>2015</pubdate>
  <volume>3</volume>
  <fpage>71</fpage>
</bibl>

</refgrp>
} 

\section{Appendix}
%

\section{Data}
\label{supp-sec:data}
We used 9 datasets for classification and regression tasks respectively (\cref{tab:dataset} in the paper). Here we present a more detailed description of the data sources and preprocessing.

\subsection{Data Preprocessing}
We downloaded the original data files from UCI Machine Learning Repository~\cite{Dua:2017} and Lu\a'is Torgo's personal website\footnote{\url{https://www.dcc.fc.up.pt/~ltorgo/Regression/DataSets.html}}. In addition, we conducted further data cleaning before the experiments\footnote{\url{https://github.com/peterfennell/S3D/blob/paper-replication/data/download-datasets.ipynb}}:

\begin{itemize}
\item \emph{Classification}: \textit{(1)} For breast cancer (original), We dropped sample code number columns and removed samples with missing values (all in feature ``Bare Nuclei '').  \textit{(2)} For parkinsons, we dropped ASCII subject name and recording number.
\item \emph{Regression}: \textit{(1)} For appliances energy use, we dropped date columns; \textit{(2)} For residential building, we dropped sales column when the target is costs and vice versa; \textit{(3)} For parkinsons, we dropped columns subject number, age, sex, and test time, which are individual subject information not used for prediction in the original paper~\cite{parkinsons-regression}. Finally, we note that log transformation of the target values were applied to appliances energy use and both residential building to reduce skewness. Specifically, we used $log_{2}(x+1)$ to avoid the logarithm of zero values.
\end{itemize}

\noindent The five remaining  human behavior datasets are thoroughly described in \cref{sec:results} of the paper.

\subsection{Cross Validation}
\label{supp-sec:cv}
To evaluate prediction performance, we applied \emph{nested cross validation (CV)} on all datasets (\cref{fig:crossval}). Specifically, we split each dataset into \emph{five} equal-size folds (a.k.a., outer folds; 1 to 5 in \cref{fig:crossval}). Each of the five was picked as test set (i.e., held out from the training and validation process). Given each outer test set (fold 1 in the example), we conducted \emph{inner CV} to find out the best hyperparameters for each predictive model based on the average performance on all four folds (fold 2, 3, 4, and 5.) During \emph{inner CV}, we conducted 4-fold CV, where each of the four folds were held out as validation set (fold 2) and the rest (fold 3, 4, and 5) were inputs to the model. Different sets of \emph{outer CV} may produce different ``best'' hyperparameters. The final prediction performance is the average value of those on each test set.

\begin{figure}[!h]
\centering
  \includegraphics[trim={0 5cm 0 2.1cm},clip,scale=0.45]{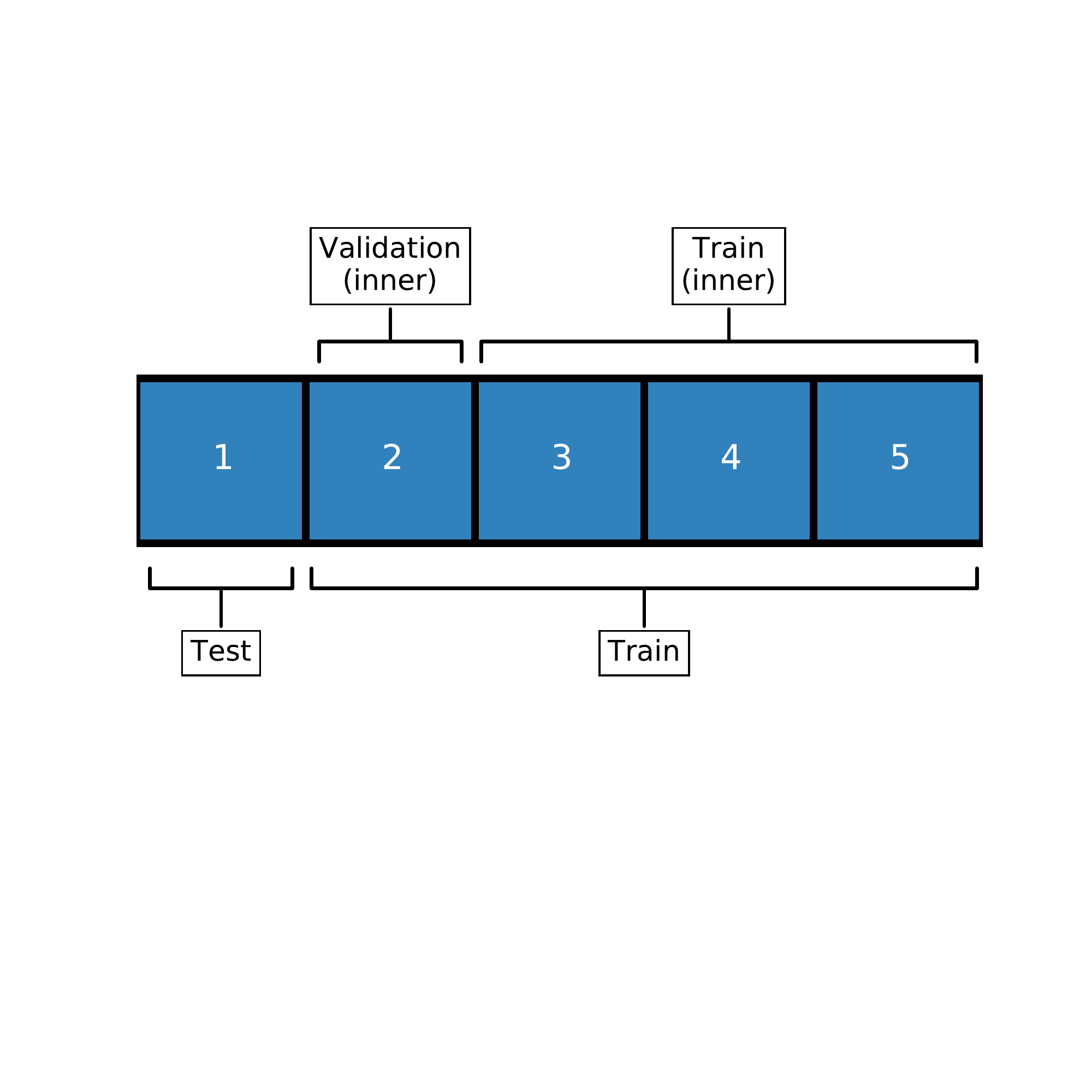}
    \caption{Demonstration of nested cross validation for one iteration.}
	\label{fig:crossval}
\end{figure}

\subsection{Data Standardization}
When applying logistic regressions~\cite[page 63]{Hastie2009} and linear support vector machine (SVM)~\cite{hsu2003practical}, we standardized all features (i.e., subtracting the mean and scaling to unit variance). For all regression data, regardless of regressors, we standardized all columns (i.e., features as well as targets). Note that in both cases, we transformed values in the test set based on the mean and variance of values in the training set.

\section{Visualizing Data with S3D}
\label{supp-sec:visualization}
One of the strengths of S3D is its ability to create visualizations of learned models for data exploration. In this section, we provide a detailed description of S3D visualizations of \emph{Stack Exchange} and \emph{Digg} data, starting from the simplest models that partition the data on the most important feature, to the increasingly more complex models that partition the feature space of several most important features.

\subsection{Stack Exchange}
The first important feature in Stack Exchange data, \emph{the number of answers before}, reflects users' experience. Intuitively, the more answers a user has written in the past, the more likely that this user is to be active, knowledgeable, and experienced; therefore, the more likely the current answer is to be accepted as best answer by the asker. This is indeed reflected by the model learned by S3D, shown in~\cref{fig:stackexchange_1}. Evidently, acceptance probability increases monotonically with user experience (~\cref{fig:stackexchange_1}(a)). On the other hand, most users provide fewer than 1,000 answers. While exceptional activity leads to remarkable acceptance rate, these users are rare in the community (~\cref{fig:stackexchange_1}(b)).

\begin{figure}[!h]
\centering
\begin{subfigure}{.5\textwidth}
  \centering
  \includegraphics[width=.8\linewidth]{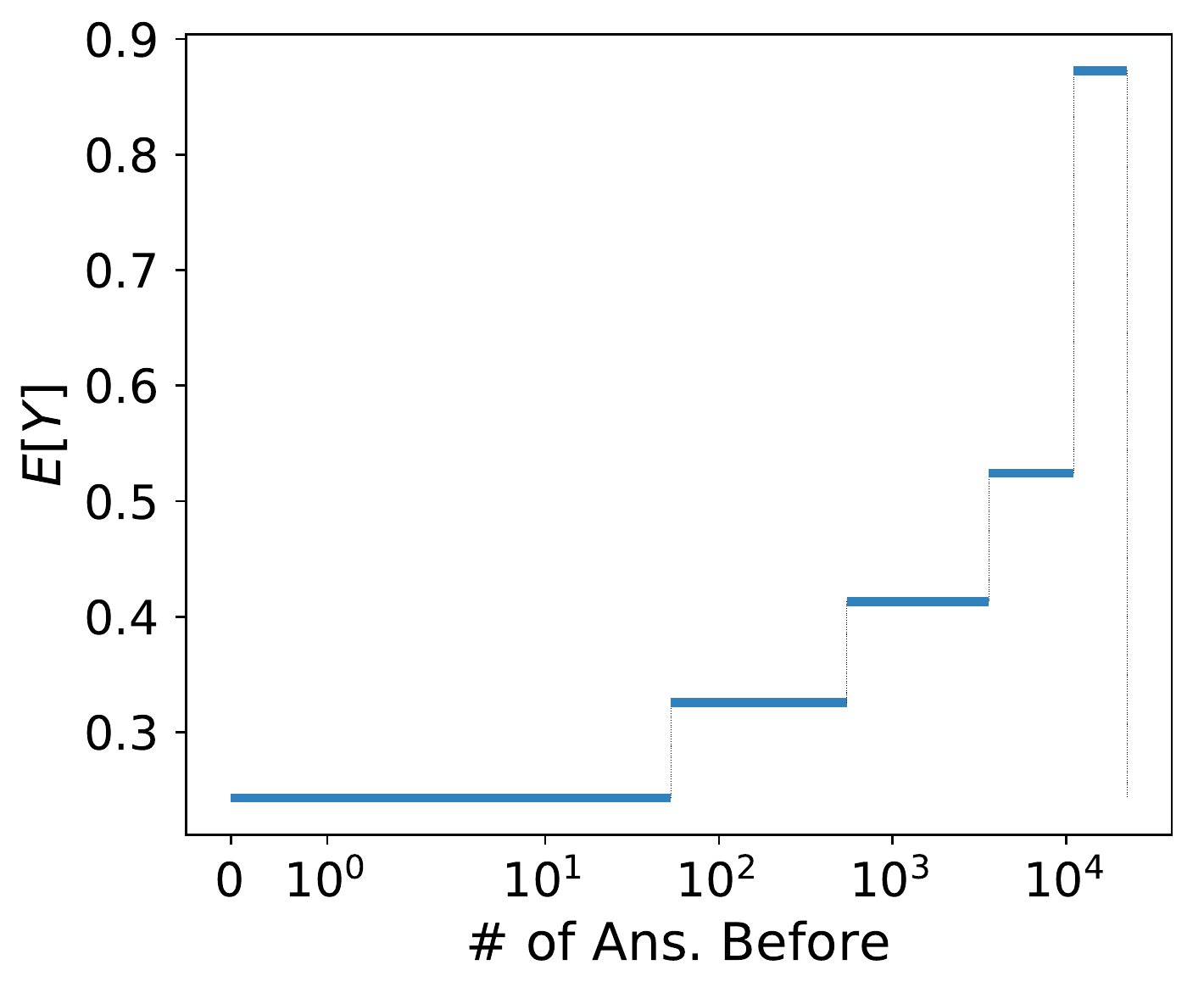}
  \caption{Expected acceptance probability}
  \label{fig:ey_stackoverflow_1}
\end{subfigure}%
\begin{subfigure}{.5\textwidth}
  \centering
  \includegraphics[width=.8\linewidth]{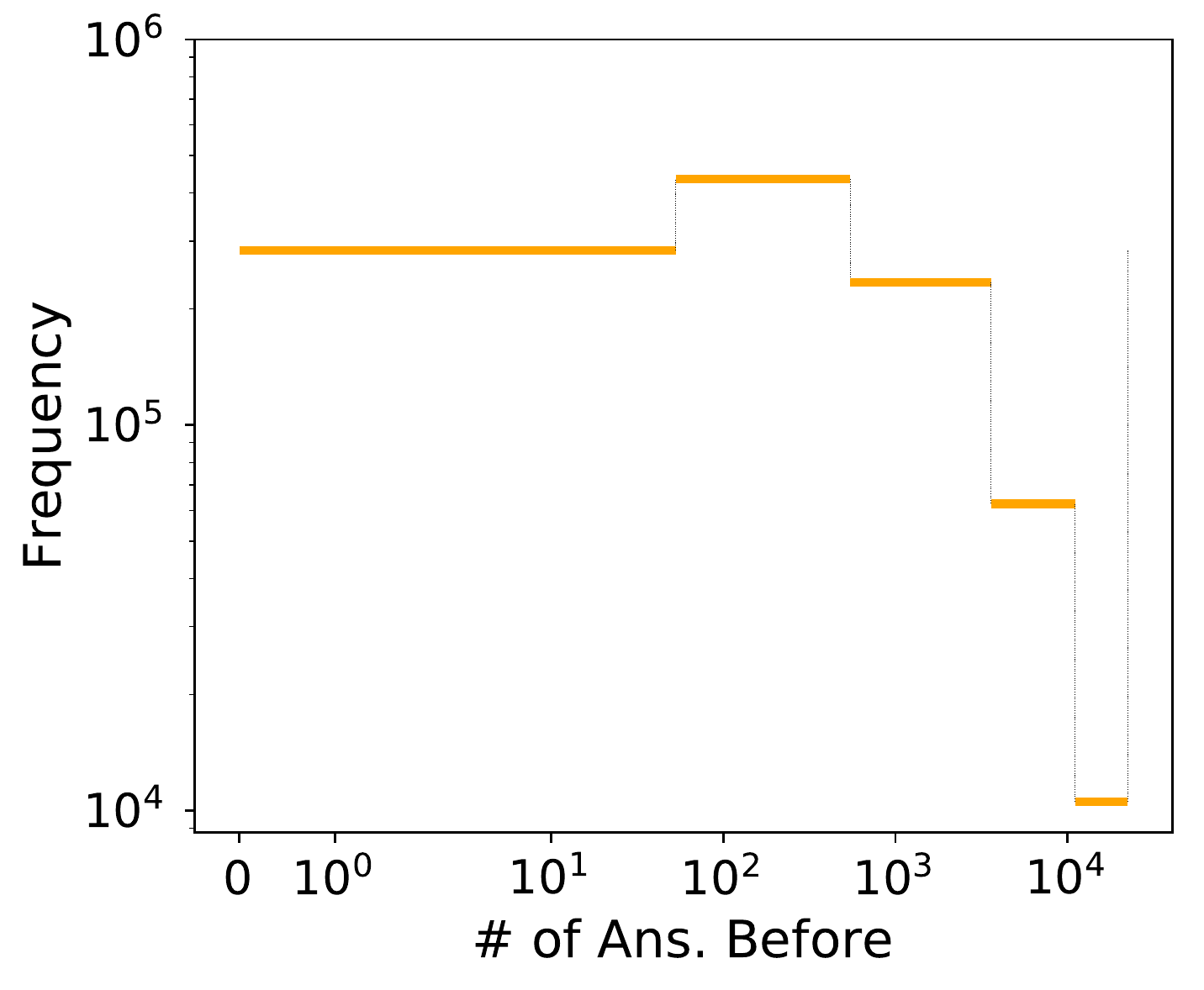}
  \caption{Frequency}
  \label{fig:freq_stackoverflow_1}
\end{subfigure}
\caption{Model of Stack Exchange learned by S3D showing the partition for the first important feature, \emph{the number of answers before}. Each horizontal line in (a) represents average acceptance probability of answers in each bin of the partition. For example, when the number of answers written by the user before the current answer is greater than $10^4$ (i.e., the last bin), there is an acceptance probability as high as 0.87; however, relatively few (10,557) samples are in that bin (b), among  the lowest of all  bins in the partition.}
\label{fig:stackexchange_1}
\end{figure}

In the second step, S3D included \emph{signup percentile} as an additional feature to explain acceptance probability (\cref{fig:stackexchange_2}). This feature represents user's rank by length of tenure among all other answerers. Overall, we can see an increasing trend of acceptance probability from the bottom left corner to top right, implying a positive relationship where more senior and experienced users are  more likely to get their answers accepted as best answers. Indeed, we can see from \cref{fig:freq_stackoverflow_2} that senior users are usually those who have written more answers than newcomers. In the mean time, acceptance probability goes down when signup percentile is lower, given the number of answers beyond 3,587. This interesting pattern indicates that answers provided by highly productive users, newcomers, rather than seniors, are more likely to get their answers accepted as best answers.

\begin{figure}[!h]
\centering
\begin{subfigure}{.5\textwidth}
  \centering
  \includegraphics[width=.8\linewidth]{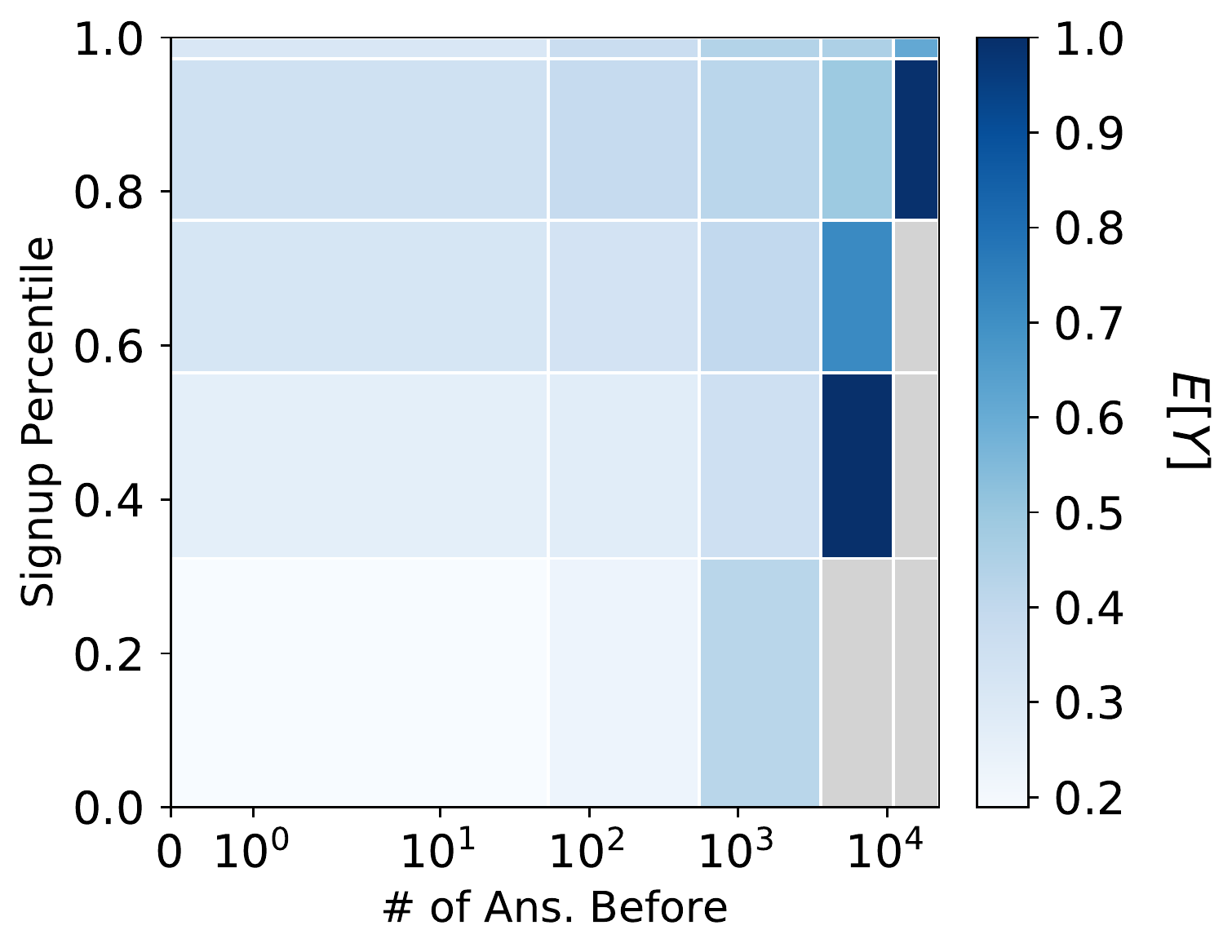}
  \caption{Expected acceptance probability}
  \label{fig:ey_stackoverflow_2}
\end{subfigure}%
\begin{subfigure}{.5\textwidth}
  \centering
  \includegraphics[width=.8\linewidth]{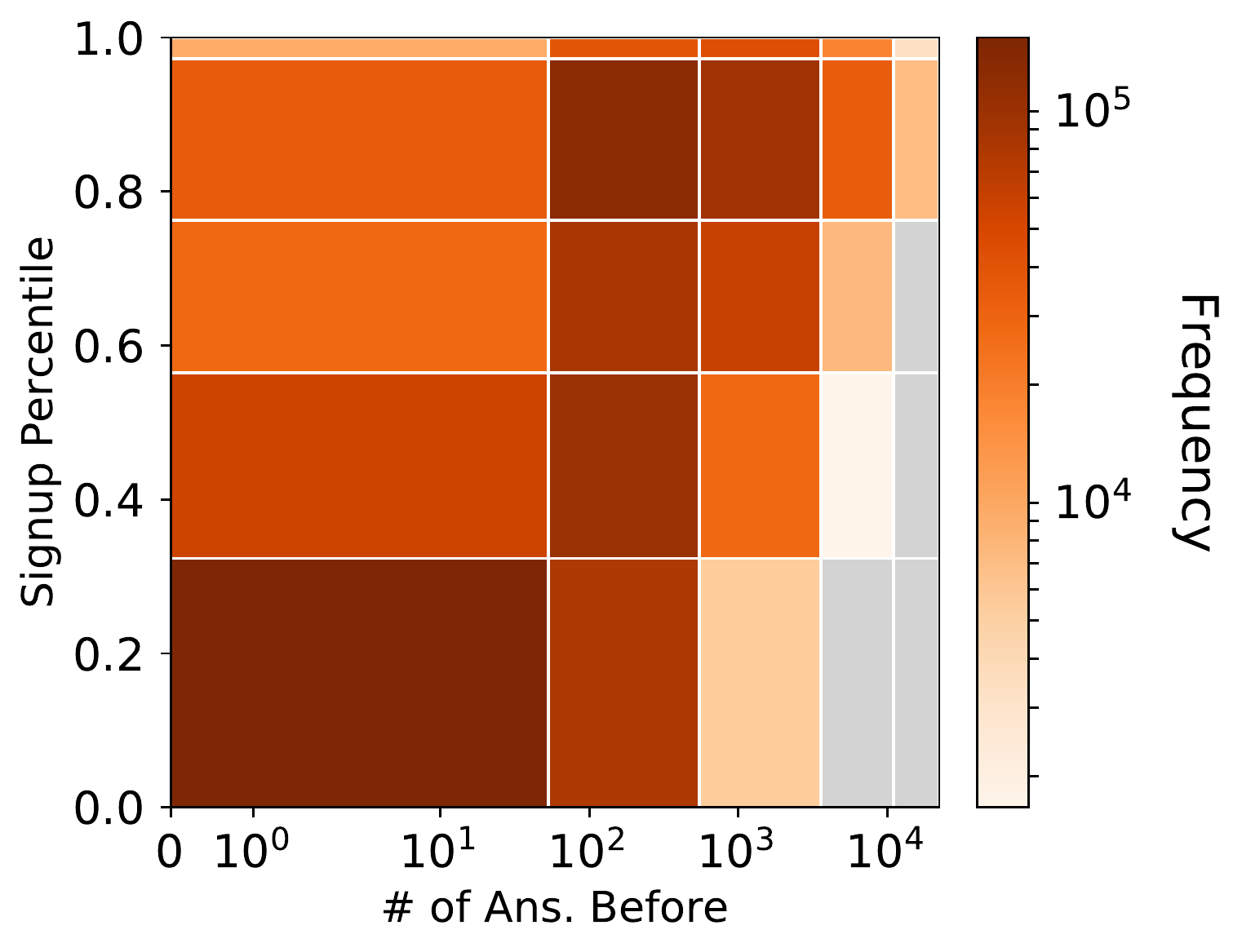}
  \caption{Frequency}
  \label{fig:freq_stackoverflow_2}
\end{subfigure}
\caption{S3D model of Stack Exchange data based on two features, \emph{the number of answers before} and \emph{signup percentile}. Each cell line represents a partition of data based on both features, e.g., when the number of answers written before the current answer ranges from 3,587 to 11,033 and signup percentile is from 0.3 to 0.58, there is an acceptance probability as high as \emph{1}, although the frequency of data samples in this bin is \emph{1,666}.}
\label{fig:stackexchange_2}
\end{figure}

\begin{figure}[!h]
  \centering
  \begin{subfigure}{\linewidth}
    \centering
    \includegraphics[width=\linewidth]{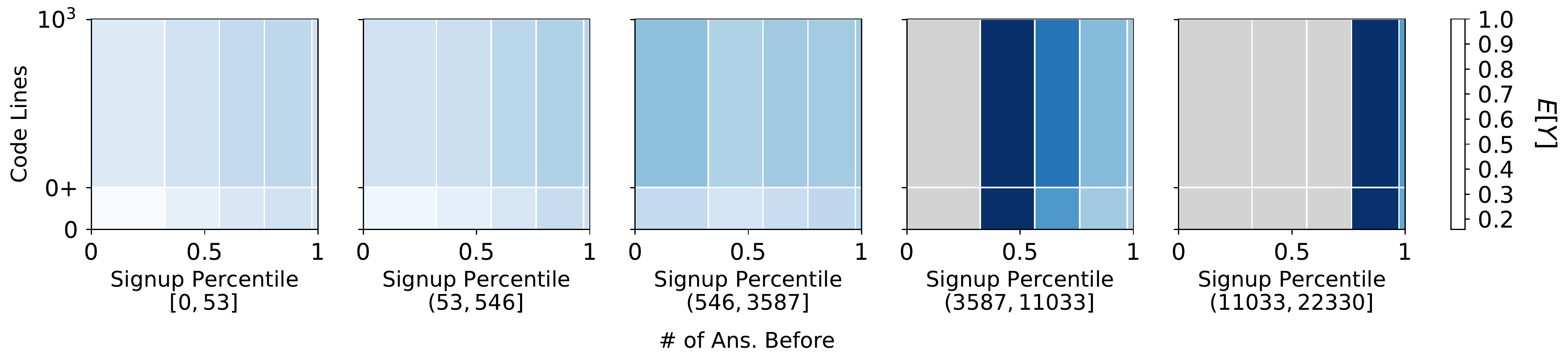}
    \caption{Expected acceptance probability}
    \label{fig:ey_stackoverflow_3}
  \end{subfigure}
  \begin{subfigure}{\linewidth}
    \centering
    \includegraphics[width=\linewidth]{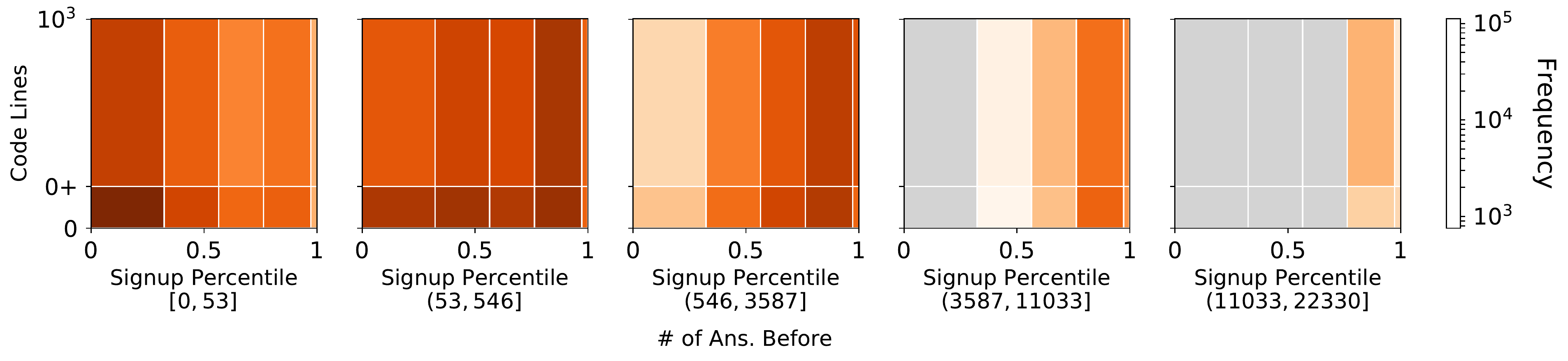}
    \caption{Frequency}
    \label{fig:freq_stackoverflow_3}
  \end{subfigure}
  \caption{S3D model for Stack Exchange data with three important features: \emph{the number of answers before}, \emph{signup percentile}, and \emph{code lines}. The interpretation is the same as \cref{fig:stackexchange_2}.}
  \label{fig:stackexchange_3}
\end{figure}
  In~\cref{fig:stackexchange_3} we proceed to look at models that include the third important feature: \emph{code lines}.
Each plot in~\cref{fig:stackexchange_3}(a) shows a bin with respect to the first important feature (number of answers written before this). The range of the bin is shown just under each plot. Note that in each plot there are two zeros on the y-axis. We manually expanded the range $[0, 0]$ from a line to a band to make the binning more clear. In fact, S3D learns from the data that answers that include programming code have a better chance of being accepted (i.e., the color in \cref{fig:ey_stackoverflow_3} is \emph{less blue} when the number of code lines is above zero), although most answers do not contain code (\cref{fig:freq_stackoverflow_3}). The model with four features can be found in the main paper in \cref{fig:interpretation}.

\begin{figure}[!h]
  \centering
  \begin{subfigure}{\linewidth}
    \centering
    \includegraphics[width=0.9\linewidth]{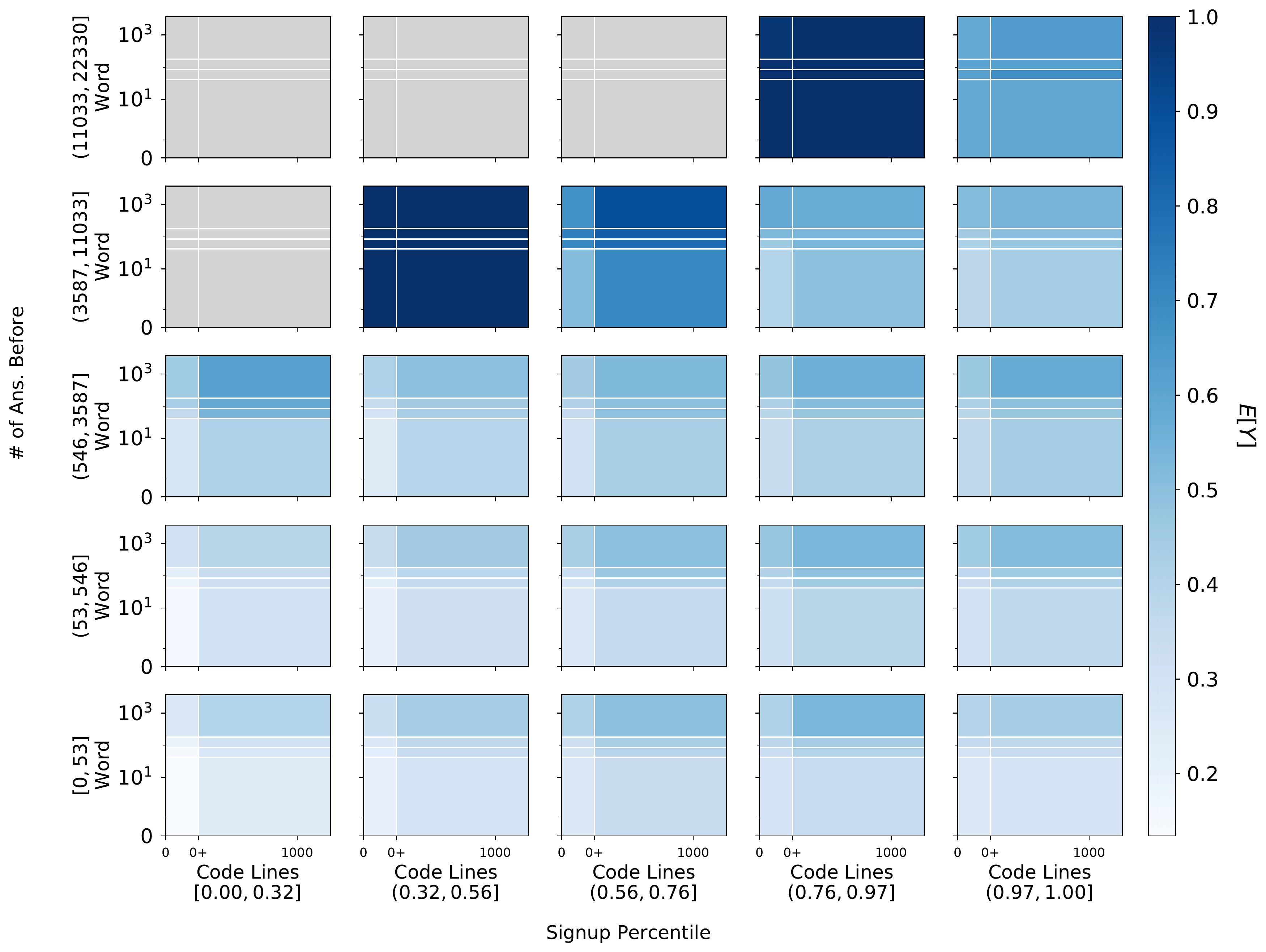}
    \caption{Expected acceptance probability}
    \label{fig:ey_stackexchange_4}
  \end{subfigure}
  \begin{subfigure}{\linewidth}
    \centering
    \includegraphics[width=0.9\linewidth]{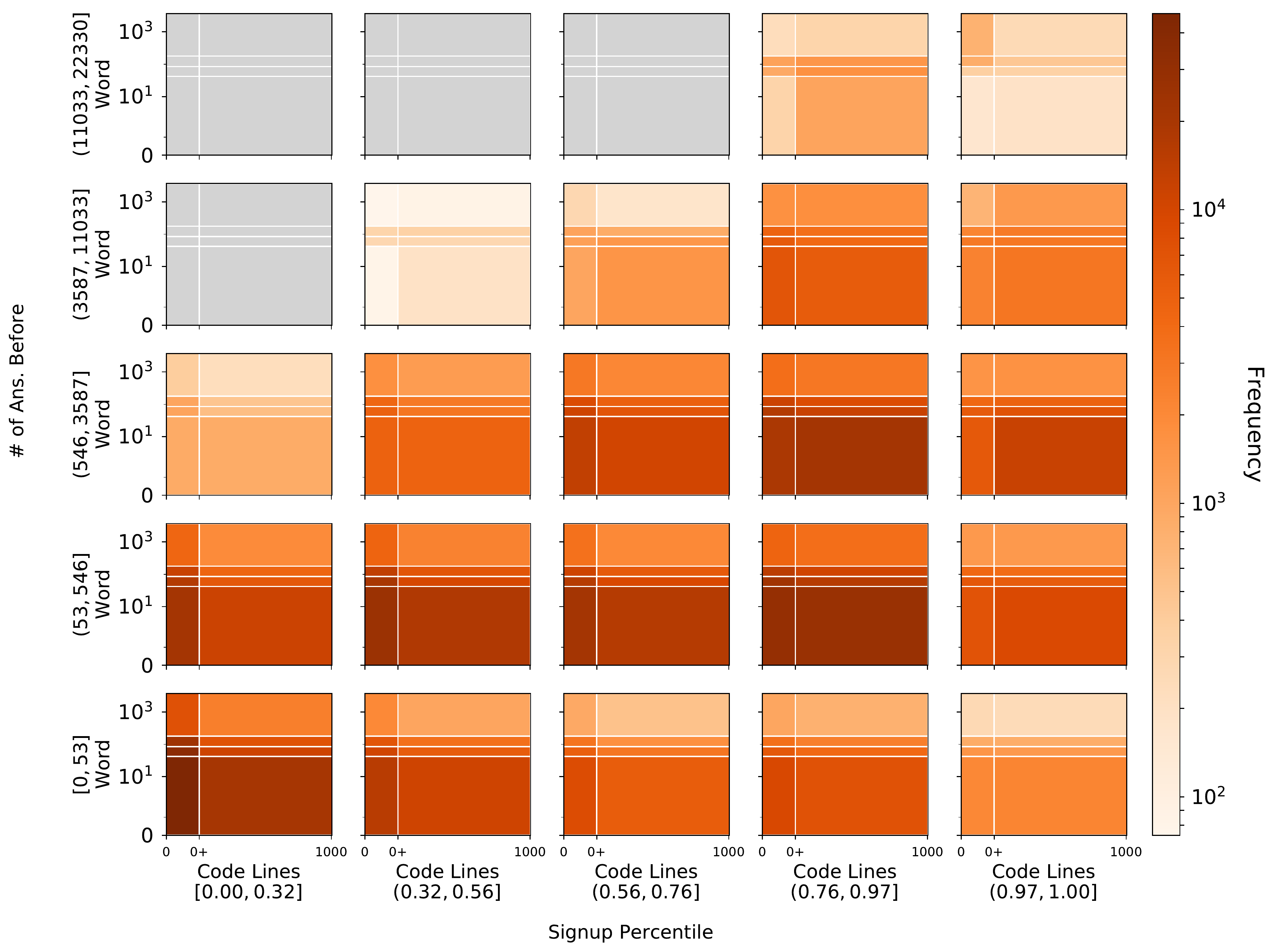}
    \caption{Frequency}
    \label{fig:freq_stackexchange_4}
  \end{subfigure}
  \caption{S3D model of Stack Exchange data with four important features: \emph{the number of answers before}, \emph{signup percentile}, \emph{code lines}, and \emph{words}, showing probability of acceptance (top) and number of samples in each bin (bottom).}
  \label{fig:stackexchange_4}
\end{figure}

Finally,~\cref{fig:ey_stackexchange_4} shows the S3D model with the fourth important feature, \textit{word}, which is the number of words in an answer. Generally, the longer---and presumably the more informative---the answer, the more likely it is to get accepted. Meanwhile, the distribution of words among answers mostly stayed in the bottom half (i.e.,  below 178 word counts) as shown in~\cref{fig:freq_stackexchange_4}. See~\cref{sec:model_analysis} for an in-depth discussion.

\subsection{Digg}
The first important feature identified by S3D in the Digg data is \emph{user activity} (\cref{fig:digg_1}). Recall that the target variable is whether or not a user will ``digg'' a story  a story following an exposure by a friend. Digging a story is similar to retweeting a post on Twitter, hence, we generically refer to it as ``adopting'' a story or a meme. While, intuitively, more active users are more likely to ``digg'' stories, the increment as shown in \cref{fig:ey_digg_1} is generally very small, implying the existence of a more complex behavioral mechanism of meme  adoption.

\begin{figure}[!h]
\centering
\begin{subfigure}{.5\textwidth}
  \centering
  \includegraphics[width=.8\linewidth]{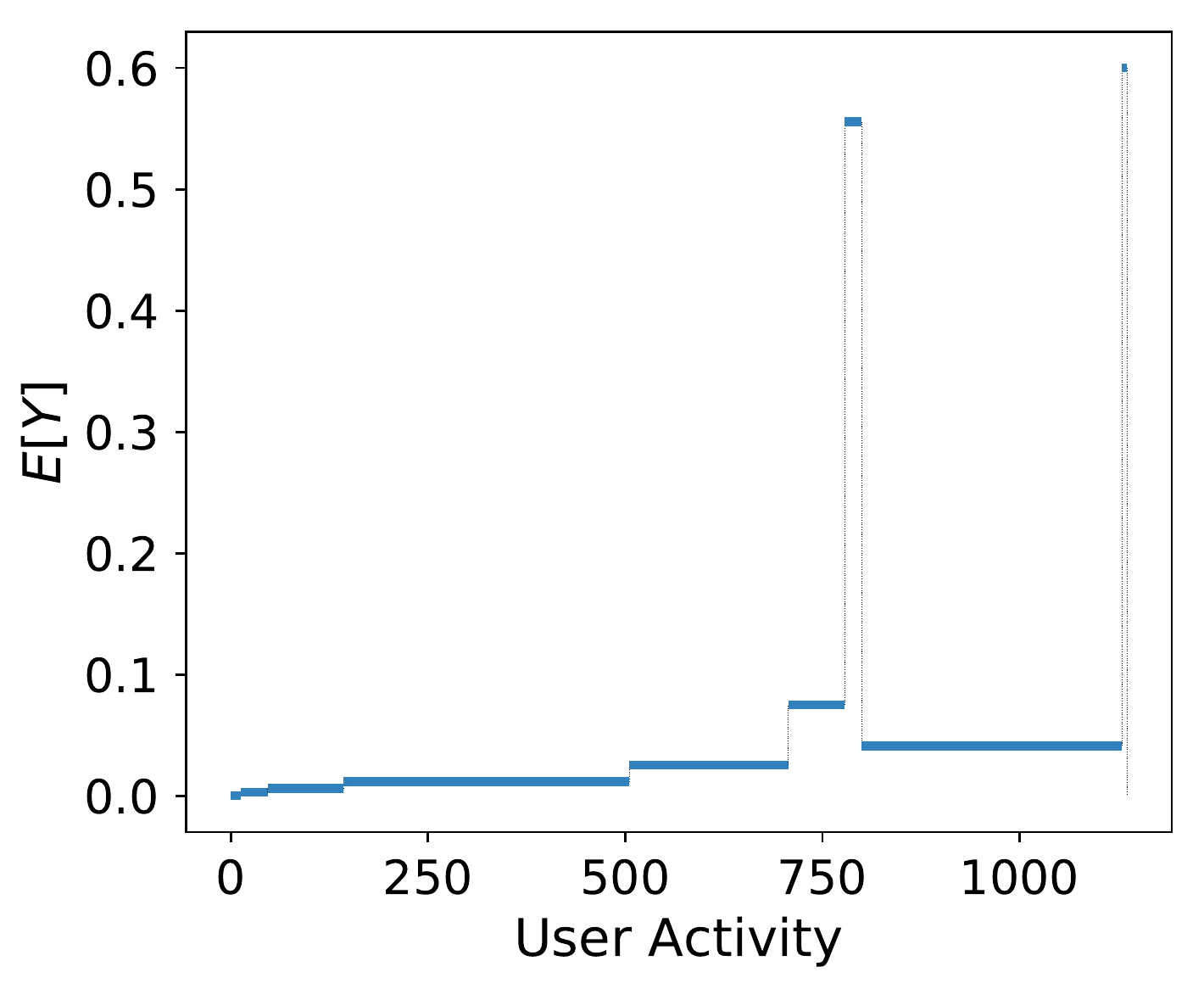}
  \caption{Expected acceptance probability}
  \label{fig:ey_digg_1}
\end{subfigure}%
\begin{subfigure}{.5\textwidth}
  \centering
  \includegraphics[width=.8\linewidth]{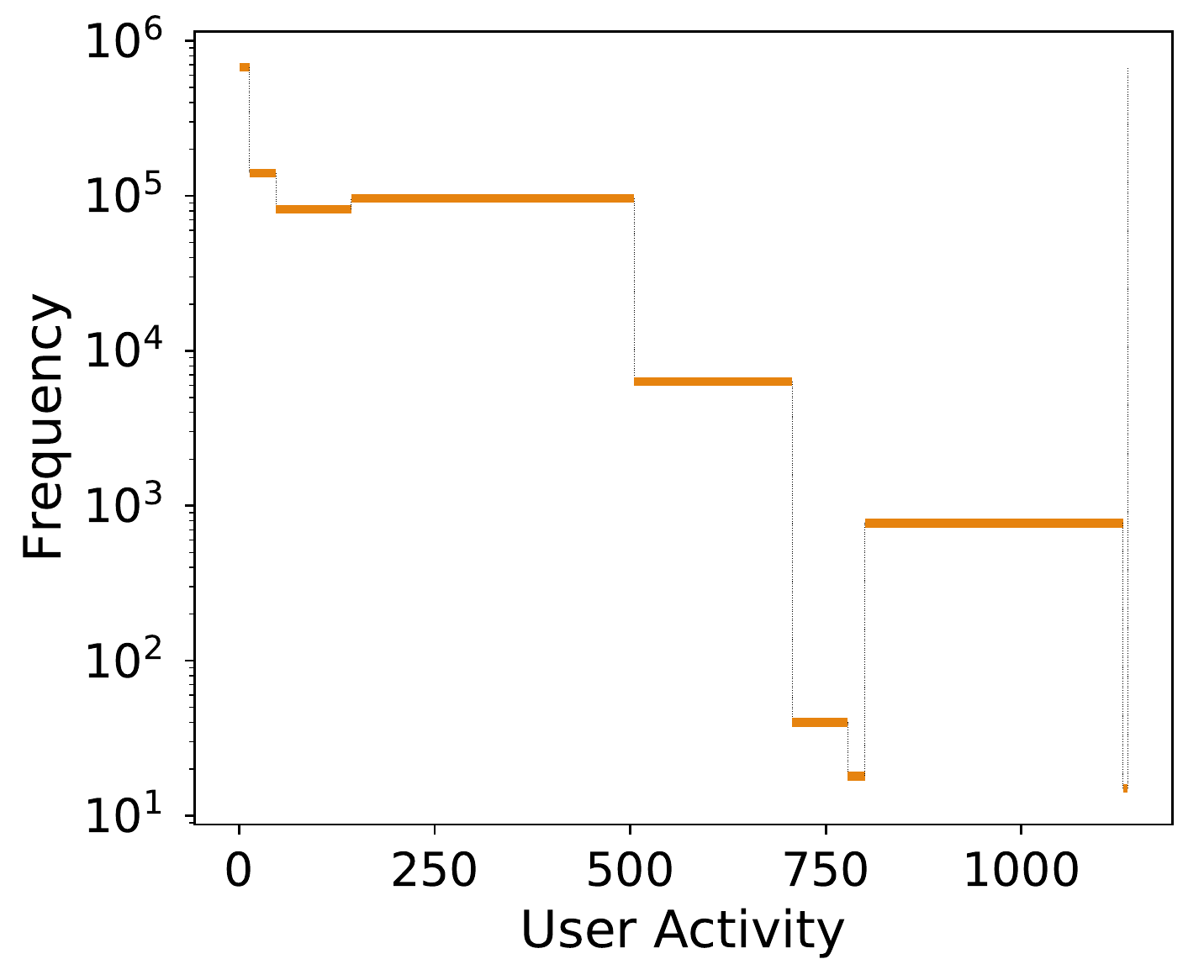}
  \caption{Frequency}
  \label{fig:freq_digg_1}
\end{subfigure}
\caption{S3D model of Digg data, showing the partition of based on the first important feature \emph{user activity}.}
\label{fig:digg_1}
\end{figure}

In the next step, S3D finds that \emph{information received by users} is an important feature (\cref{fig:digg_2}), with which S3D discovers a more fine-grained partition of data. This feature describes the user's information load, that is the number of stories ``dugg'' or recommended by friends. Probability to ``digg'' increases when going from top left to bottom right. Indeed, active users who have lower information load are more likely to ``digg'' the stories they receive. Nonetheless, most users are not active but receive a huge load of information from their neighbors (\cref{fig:freq_digg_2}). See \cref{fig:digg} in the paper for the model with three features.

\begin{figure}[!h]
\centering
\begin{subfigure}{.5\textwidth}
  \centering
  \includegraphics[width=.8\linewidth]{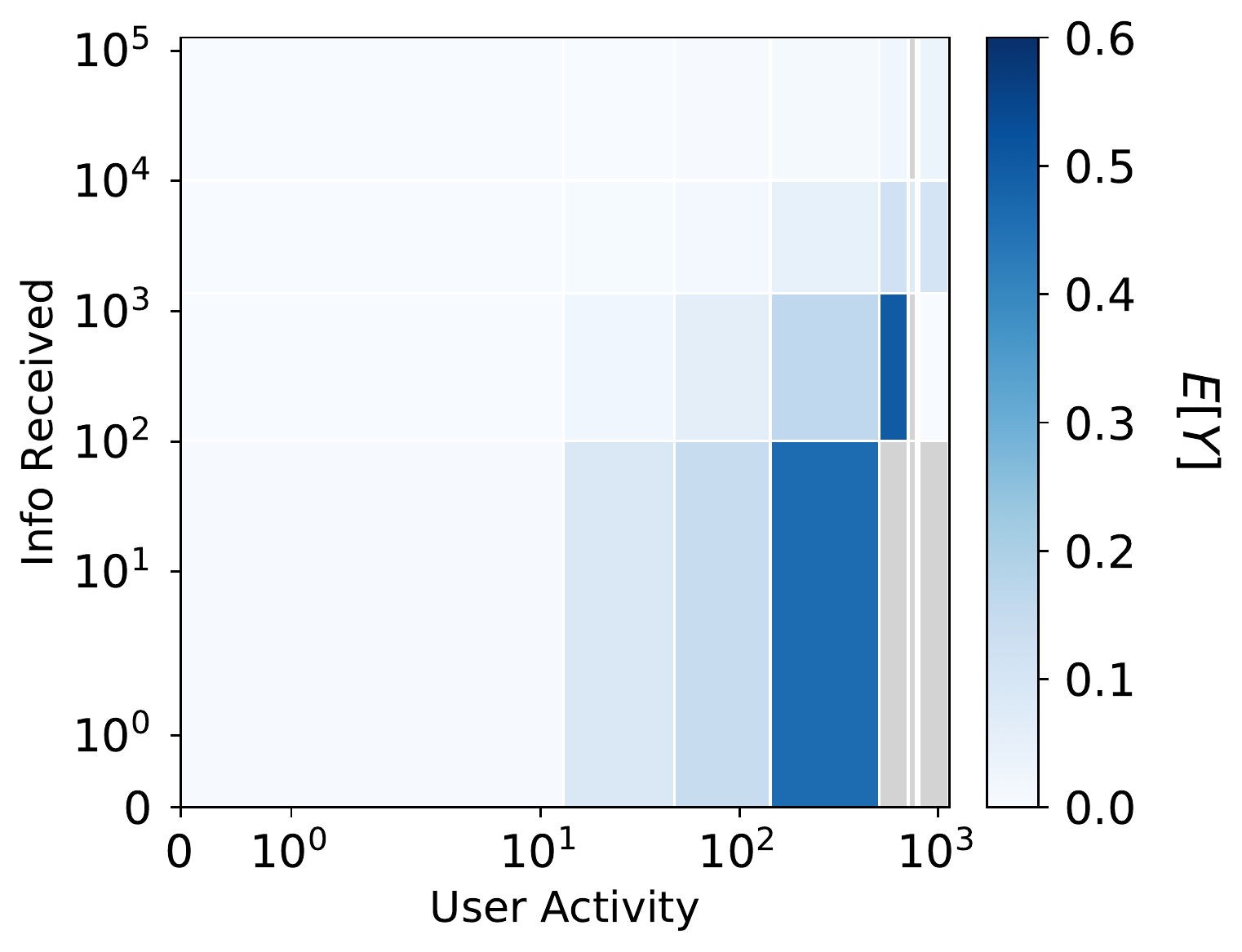}
  \caption{Expected acceptance probability}
  \label{fig:ey_digg_2}
\end{subfigure}%
\begin{subfigure}{.5\textwidth}
  \centering
  \includegraphics[width=.8\linewidth]{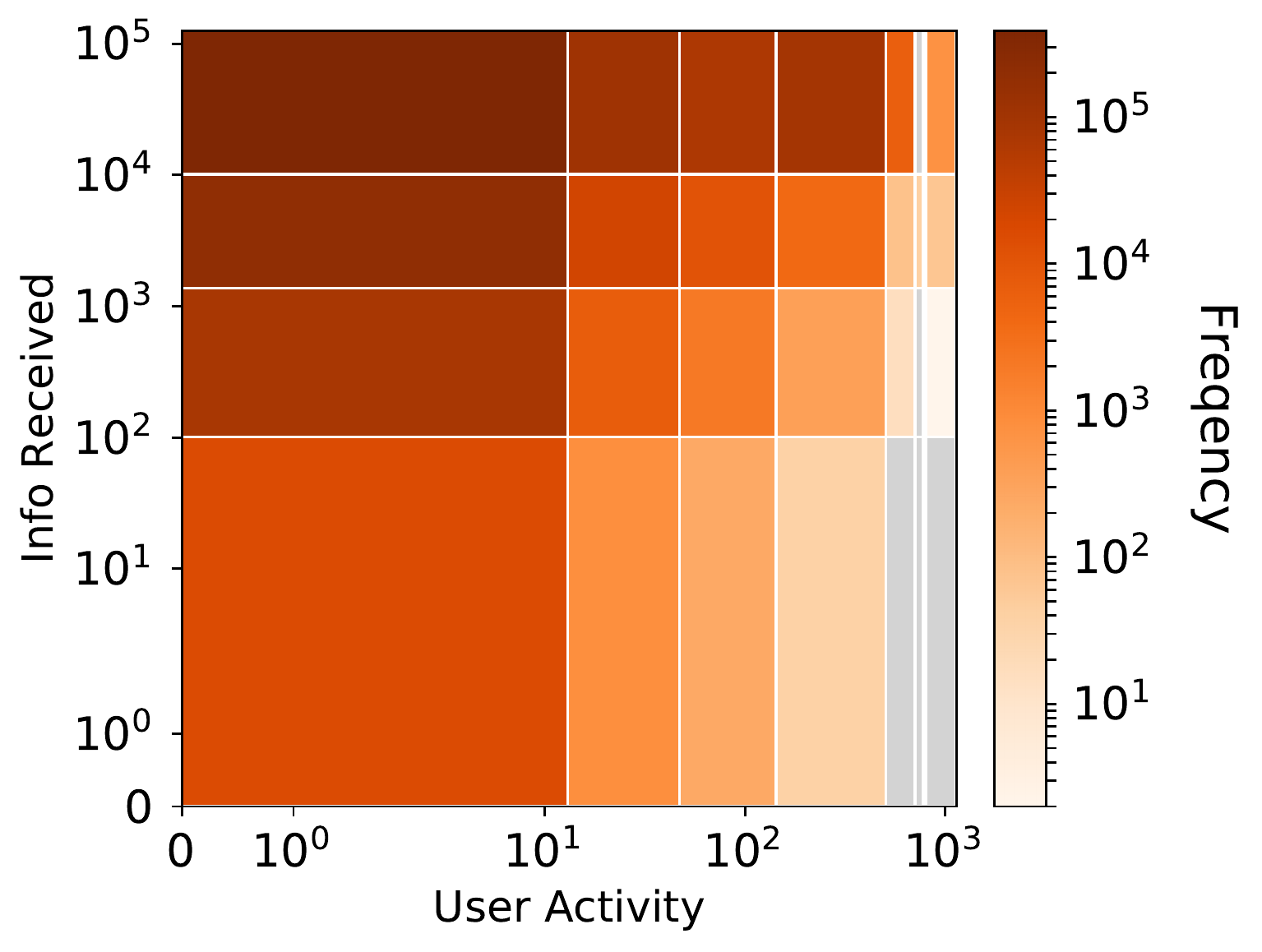}
  \caption{Frequency}
  \label{fig:freq_digg_2}
\end{subfigure}
\caption{S3D model for Digg with two feature, \emph{user activity} and \emph{information received}.}
\label{fig:digg_2}
\end{figure}

The last feature that S3D picked is \textit{the current popularity of this meme}, which measures how ``viral'' a story is (\cref{fig:ey_digg_3}). See \cref{sec:model_analysis} for a more detailed description on the learned model. Overall, users tend to ``digg '' a popular story, implying the \textit{Matthew effect} on diffusion of Digg stories.
In~\cref{fig:freq_digg_3}, it is interesting to see that most users are exposed to viral stories.

\begin{figure}[!h]
  \centering
  \begin{subfigure}{\linewidth}
    \centering
    \includegraphics[width=\linewidth]{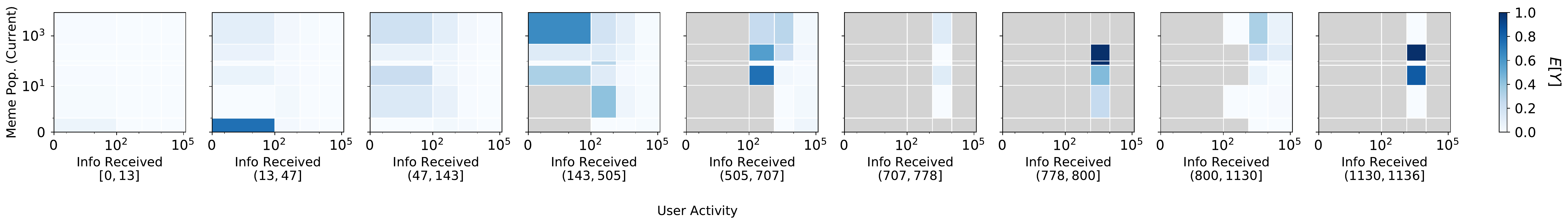}
    \caption{Expected acceptance probability}
    \label{fig:ey_digg_3}
  \end{subfigure}
  \begin{subfigure}{\linewidth}
    \centering
    \includegraphics[width=\linewidth]{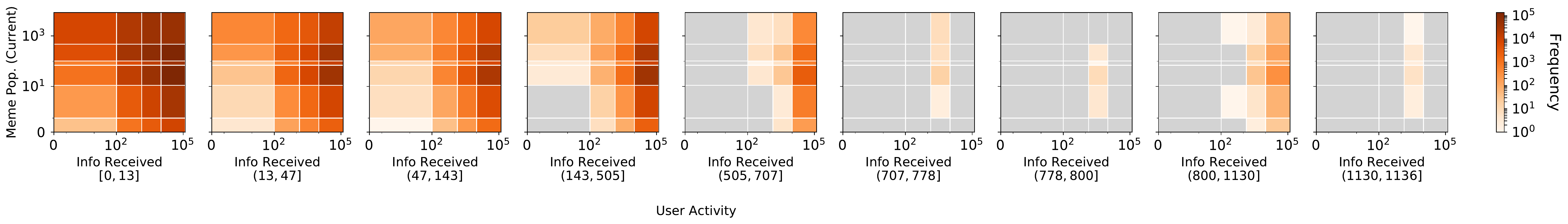}
    \caption{Frequency}
    \label{fig:freq_digg_3}
  \end{subfigure}
  \caption{Full S3D model for Digg data with three important features: \emph{user activity}, \emph{information received}, and \emph{meme current popularity}.}
  \label{fig:digg_3}
\end{figure}

\end{backmatter}
\end{document}